\documentclass[preprint,12pt]{elsarticle}

\usepackage{latexsym,amsmath,amssymb,xspace,comment,graphicx, url, color}
\usepackage{comment}
\usepackage[numbers]{natbib}
\usepackage{algorithm}
\usepackage{algpseudocode}
\usepackage{here}

\journal{}

\begin{document}

\begin{frontmatter}



\title{A table of short-period Tausworthe generators for Markov chain quasi-Monte Carlo}


\author[harase]{Shin Harase\corref{cor1}}
\ead{harase@fc.ritsumei.ac.jp}
\address[harase]{College of Science and Engineering, Ritsumeikan University, 1-1-1 Nojihigashi, Kusatsu, Shiga, 525-8577, Japan.}
\cortext[cor1]{Corresponding author}

\begin{abstract}
We consider the problem of estimating 
expectations by using Markov chain Monte Carlo methods 
and improving the accuracy 
by replacing IID uniform random points 
with quasi-Monte Carlo (QMC) points. 
Recently, it has been shown that Markov chain QMC remains consistent 
when the driving sequences are completely uniformly distributed (CUD).  
However, the definition of CUD sequences is not constructive, 
so an implementation method using short-period Tausworthe
generators (i.e., linear feedback shift register generators over the two-element
field) that approximate CUD sequences has been proposed. 
In this paper, we conduct an exhaustive search of 
short-period Tausworthe generators for Markov chain QMC
in terms of the $t$-value, which is a criterion of uniformity widely 
used in the study of QMC methods. 
We provide a parameter table of Tausworthe generators 
and show the effectiveness in numerical examples using Gibbs sampling. 
\end{abstract}

\begin{keyword}
Pseudorandom number generation \sep Quasi-Monte Carlo \sep  Markov chain Monte Carlo 
\sep Polynomial lattice point set \sep Continued fraction expansion

\MSC[2010] 65C10 \sep 11K45
\end{keyword}

\end{frontmatter}

\newtheorem{theorem}{Theorem}
\newtheorem{lemma}[theorem]{Lemma}
\newtheorem{corollary}[theorem]{Corollary}
\newdefinition{definition}{Definition}
\newdefinition{remark}{Remark}
\newproof{proof}{Proof}\newtheorem{proposition}[theorem]{Proposition}

\section{Introduction}

We consider the problem of estimating the expectation $E_{\pi} [f(\mathbf{X})]$ 
by using Markov chain Monte Carlo (MCMC) methods 
for a target distribution $\pi$ and some function $f$. 
For this problem, we want to improve the accuracy  
by replacing independent and identically distributed (IID) 
uniform random points with quasi-Monte Carlo (QMC) points. 
However, typical QMC points (e.g., Sobol', Faure, and Niederreiter--Xing) are not applicable in general. 
Motivated by a simulation study by Liao \cite{doi:10.1080/10618600.1998.10474775}, 
Owen and Tribble \cite{MR2168266} 
and Chen et al.~\cite{MR2816335} proved that Markov chain QMC remains consistent 
when the driving sequences are completely uniformly distributed (CUD). 
Here, a sequence $u_0, u_1, u_2,$ $\ldots$ $\in [0, 1)$ is said to be CUD 
if overlapping $s$-blocks 
$(u_i, u_{i+1},$ $\ldots, u_{i+s-1})$, $i = 0, 1, 2, \ldots$, are uniformly distributed for every dimension $s \geq 1$. 

Levin \cite{MR1731474} provided several constructions for CUD sequences, 
but they are not convenient to implement. 
Instead, to construct CUD sequences approximately, 
Tribble and Owen \cite{MR2426105} and 
Tribble \cite{MR2710331} proposed an implementation method 
using short-period linear congruential 
and Tausworthe generators (i.e., linear feedback shift register generators 
over the two-element field $\mathbb{F}_2 := \{ 0, 1\}$) that run 
for the entire period. 
Chen et al.~\cite{MR3173841} implemented   
short-period Tausworthe generators optimized in terms of the equidistribution property, 
which is a coarse criterion used in the area of pseudorandom number generation 
(see \cite[\S8.1]{ChenThesis} for the complete parameter table). 
In the theory of $(t, m, s)$-nets and $(t, s)$-sequences, 
the $t$-value is a central criterion of uniformity. 
In fact, typical QMC points (e.g., Sobol', Faure, and Niederreiter--Xing) 
are optimized in terms of the $t$-value (see \cite{MR1172997,MR2683394}). 

The aim of this paper is to conduct an exhaustive search of 
short-period Tausworthe generators 
for Markov chain QMC in terms of the $t$-value and 
to provide a parameter table of Tausworthe generators. 
It is known that Tausworthe generators can be viewed 
as polynomial Korobov lattice point sets with a denominator polynomial $p(x)$ 
and a numerator polynomial $q(x)$ over $\mathbb{F}_2$ 
(e.g., see \cite{MR2723077,MR1978160}). 
For dimension $s = 2$, there is a connection between the $t$-value and continued fraction expansions, 
that is, the $t$-value is optimal (i.e., the $t$-value is zero) 
if and only if the partial quotients in the continued fraction of $q(x)/p(x)$ are all of degree one. 
To satisfy the definition of CUD sequences approximately, 
we want to search for parameters $(p(x), q(x))$ whose $t$-values are optimal for $s = 2$ 
and as small as possible for $s \geq 3$. 
As a previous study, in 1993, Tezuka and Fushimi \cite{MR1160278} 
proposed an algorithm to search for such parameters 
using a polynomial analogue of Fibonacci numbers 
from the viewpoint of continued fraction expansions. 
Thus, we refine their algorithm on modern computers, 
and conduct an exhaustive search again. 
In addition, we report numerical examples using Gibbs sampling 
in which the resulting QMC point sets perform better 
than the existing point sets developed by Chen et al.~\cite{MR3173841}. 

One might consider searching for parameters $(p(x), q(x))$ 
with $t$-value zero for $s = 3$. 
Kajiura et al.~\cite{MR3807854} 
proved that there exists no maximal-period Tausworthe generator with this property. 

The remainder of this paper is organized as follows: 
In Section~\ref{sec:preliminaries}, 
we briefly recall the definition of CUD sequences, 
Tausworthe generators, and the $t$-value and equidistribution property. 
 Section~\ref{sec:main} is devoted to our main results: 
we describe an exhaustive search algorithm 
and provide a table of short-period Tausworthe generators for Markov chain QMC. 
We also compare our new generators with existing generators 
developed by Chen et al.~\cite{MR3173841} 
in terms of the $t$-value and equidistribution property. 
In Section~\ref{sec:example}, we present numerical examples 
using Gibbs sampling. 
In Section~\ref{sec:conclusion}, we conclude this paper.

\section{Preliminaries} \label{sec:preliminaries}
We refer the reader to \cite{MR1172997,MR2683394,MR2723077,LP2009} for general information. 
\subsection{Discrepancy and completely uniformly distributed sequences}
 
Let $P_s = \{ \mathbf{u}_0,  \mathbf{u}_1, \ldots, \mathbf{u}_{N-1} \} \subset [0, 1)^s$ 
be an $s$-dimensional point set of $N$ elements in the sense of a ``multiset". 
We recall the definition of the discrepancy as a criterion of uniformity of $P_s$. 
\begin{definition}[Discrepancy] \label{def:discrepancy}
For a point set $P_s = \{ \mathbf{u}_0,  \mathbf{u}_1, \ldots, \mathbf{u}_{N-1} \} \subset [0,1)^s$, the (\textit{star}) \textit{discrepancy} is defined as 
\[ D_N^{*s} (P_s) := \sup_{J} \left| \frac{\nu (J; P_s)}{N} - \textrm{vol}(J) \right|, \]
where the supremum is taken over every sub-interval $J = [0, t_1) \times \cdots \times [0, t_s) \subset [0,1)^s$, 
$\nu (J; P_s)$ is the number of points from $P_s$ 
that belong to $J$, and $\textrm{vol}(J) := t_1 \cdots t_s$ is the volume of $J$. 
\end{definition}
If $D_N^{*s} (P_s)$ is close to zero, we regard $P_s$ as highly uniformly distributed. 

Next, we define the CUD property 
for a one-dimensional infinite sequence $\{ u_i \}_{i = 0}^{\infty} \subset [0,1)$.
\begin{definition}[CUD sequences] \label{def:cud} 
A one-dimensional infinite sequence $u_0, u_1, u_2,$ $\ldots$ $\in [0, 1)$ 
is said to be \textit{completely uniformly distributed} (\textit{CUD}) 
if overlapping $s$-blocks satisfy  
\begin{eqnarray*} \label{eqn:cud}
 \lim_{N \to \infty} D_N^{*s} \left( (u_0, \ldots, u_{s-1}), (u_1, \ldots, u_{s}), \ldots, (u_{N-1}, \ldots, u_{N+s-2}) \right) = 0
 \end{eqnarray*}
for every dimension $s \geq 1$, that is, the sequence of $s$-blocks $(u_i, \ldots, u_{i+s-1}), i=0,1, \ldots$, is 
uniformly distributed in $[0,1)^s$ for every dimension $s \geq 1$. 

\end{definition}
This is one of the definitions of a random sequence from Knuth \cite{MR3077153}. 
From the viewpoint of QMC, it is desirable that $D_N^{*s}$ 
converges to zero fast if $N \to \infty$; see \cite{MR3563205,MR3275857} for details. 
As a necessary and sufficient condition of Definition~\ref{def:cud}, Chentsov \cite{CHENTSOV1967218} showed that 
non-overlapping blocks satisfy 
\[ \lim_{N \to \infty} D_N^{*s} \left((u_0, \ldots, u_{s-1}), (u_{s}, \ldots, u_{2s-1}), \ldots, (u_{s(N-1)}, \ldots, u_{Ns-1}) \right) = 0 \]
for every dimension $s \geq 1$. Thus, we use a sequence $\{ u_i \}_{i = 0}^{\infty} \subset [0,1)$ for Markov chain QMC in this order. 

\subsection{Tausworthe generators}

We recall some results of Tausworthe generators. 
Let $\mathbb{F}_2:= \{ 0, 1\}$ be the two-element field, and 
perform addition and multiplication over $\mathbb{F}_2$ (or modulo 2). 

\begin{definition}[Tausworthe generators \cite{MR0184406,MR1325871,MR1620231}] \label{def:Tausworthe}
Let $p(x) := x^m - c_1 x^{m-1} - \cdots - c_{m-1}x - c_m \in \mathbb{F}_2[x]$.
Consider the linear recurrence 
\begin{eqnarray} \label{eqn:m-sequence}
a_i  := c_1 a_{i-1} + \cdots + c_m a_{i-m} \in \mathbb{F}_2,
\end{eqnarray}
whose \textit{characteristic polynomial} is $p(x)$. Let $\sigma$ be a \textit{step size} with $0 < \sigma < 2^m-1$ and 
\begin{eqnarray} \label{eqn:Tausworthe}
u_i := \sum_{j = 0}^{w-1} a_{i \sigma +j} 2^{-j-1} \in [0,1)
\end{eqnarray}
be the \textit{output} at step $i$, 
where $w$ is the \textit{word size} of the intended machine.
If $p(x)$ is primitive, $(a_0, \ldots, a_{m-1}) \neq (0, \ldots, 0)$, and $\gcd (\sigma, 2^m-1) = 1$, 
then the sequences (\ref{eqn:m-sequence}) and (\ref{eqn:Tausworthe}) are both purely periodic with maximal period $2^m-1$. 
Assume the maximal periodicity and $\sigma \geq w$. 
A generator in such a class is called a \textit{Tausworthe generator} 
(or a \textit{linear feedback shift register generator}). 
\end{definition}

Let $N = 2^m$ and consider a sequence
\begin{eqnarray} \label{eqn:Tausworthe sequence}
u_0, u_1, \ldots, u_{N-2}, u_{N-1}= u_0, \ldots \in [0,1)
\end{eqnarray}
generated from a Tausworthe generator with the period length $N-1$. 
We consider $s$-dimensional overlapping points $\mathbf{u}_i = (u_i, \ldots, u_{i+s-1})$ for $i=0,1, \ldots, N-2$, 
that is, $\mathbf{u}_0 = (u_0, \ldots, u_{s-1}), \mathbf{u}_1 = (u_1, \ldots,  u_s), \ldots, \mathbf{u}_{N-2} = (u_{N-2}, u_0, \ldots, u_{s-2})$.  
Adding the origin $\{ \mathbf{0} \}$, 
we regard a point set
\begin{eqnarray} \label{eqn:point set}
P_s = \{ \mathbf{0} \} \cup \{ \mathbf{u}_i \}_{i = 0}^{N-2} \subset [0,1)^s
\end{eqnarray}
as a QMC point set. Note that the cardinality is $|P_s| = 2^m$.

Moreover, Tausworthe generators can be represented as a polynomial analogue of linear  congruential generators: 
\begin{eqnarray}
q(x) & := & x^{\sigma} \mod{p(x)} \label{eqn:Tausworthe1} \\
X_i(x) & := & q(x) X_{i-1}(x) \mod{p(x)} \label{eqn:Tausworthe2}\\
{X_i(x)}/{p(x)} & = & a_{i \sigma} x^{-1} + a_{i \sigma+1}x^{-2} + a_{i \sigma+2}x^{-3} + \cdots \in \mathbb{F}_2((x^{-1})) 
\label{eqn:Tausworthe3}.
\end{eqnarray}
Then, the sequence (\ref{eqn:Tausworthe}) is expressed as $u_i = \nu_w({X_i(x)}/{p(x)})$, 
where a map $\nu_w : \mathbb{F}_2((x^{-1})) \to [0,1)$ is given by $\sum_{j = j_0}^{\infty} k_j x^{-j-1} \mapsto \sum_{j = \max{\{0,  j_0 \}}}^{w-1} k_{j} 2^{-j-1}$, which is obtained by 
substituting $x=2$ into (\ref{eqn:Tausworthe3}) and 
truncating the value with the word size $w$. 
Furthermore, according to \cite[\S~5.5]{MR2723077} and \cite{MR1978160}, 
a point set $P_s$ in (\ref{eqn:point set}) can also be represented as a \textit{polynomial Korobov lattice} point set: 
\begin{eqnarray} \label{eqn:Kolobov}
 P_s = \biggl\{ \nu_w \left(\frac{h(x)}{p(x)} (1, q(x), q(x)^2, \ldots, q(x)^{s-1}) \right) \Big| \ \deg (h(x)) < m \biggl\},
\end{eqnarray}
where $m = \deg (p(x))$ and the map $\nu_w$ is applied component-wise. 
A pair of polynomials $(p(x), q(x))$ is a parameter set of $P_s$. 
Thus, to construct a point set that approximates  
CUD sequences in Definition~\ref{def:cud}, 
we want to find a pair $(p(x), q(x))$ with small discrepancies $D_N^{*s}(P_s)$ for each $s \geq 1$. 


\subsection{Criteria of uniformity}

Generally, calculating $D_N^{*s}(P_s)$ is NP-hard \cite{MR2513611}. 
A point set $P_s$ in (\ref{eqn:point set}) generated from a Tausworthe generator is a \textit{digital net}, 
so we can compute the $t$-value closely related to $D_N^{*s}(P_s)$ for $N = 2^m$.
\begin{definition}[$(t, m, s)$-nets] \label{def:(t, m, s)-net}
Let $s \geq 1$ and $0 \leq t \leq m$ be integers. 
Then, a point set $P_s$ consisting of $2^m$ points in $[0, 1)^s$ is called a \textit{$(t, m, s)$-net} (in base $2$) 
if every subinterval $E = \prod_{j = 1}^s [ {r_j}/2^{d_j}, {(r_j +1)}/2^{d_j} )$ in $[0, 1)^s$ 
with integers $d_j \geq 0$ and $0 \leq r_j < 2^{d_j}$ for $1 \leq j \leq s$ 
and of volume $2^{t-m}$ contains exactly $2^t$ points of $P_s$. 
\end{definition}
For dimension $s$, 
the smallest value $t$ for which $P_s$ is a $(t, m, s)$-net 
is called the \textit{$t$-value}. 
$D_N^{*s}(P_s) = O(2^t(\log N)^{s-1}/N)$ holds, 
where the implied constant in the $O$-notation only depends on $s$, 
so a small $t$-value is desirable. 
Thus, we want to find Tausworthe generators with pairs of polynomials 
$(p(x), q(x))$ whose $t$-values are optimal (i.e., $t = 0$) 
for $s = 2$ and as small as possible for $s \geq 3$. 
Note that all Tausworthe generators have the $t$-value zero for $s = 1$. 

Conversely, Chen et al.~\cite{MR3173841} 
used the following equidistribution property as a criterion of uniformity: 
\begin{definition}[$s$-dimensional equidistribution with $l$-bit accuracy]
For $1 \leq s \leq m$ and $1 \leq l \leq m$, 
a point set $P_s$ consisting of $2^m$ points in $[0, 1)^s$ is said to be
\textit{$s$-dimensionally equidistributed with $l$-bit accuracy}  
if we can partition the $s$-dimensional unit cube $[0,1)^s$ 
into congruent cubic boxes of volume $2^{-sl}$ 
by dividing each axis $[0,1)$ into $2^l$ intervals, 
and can obtain an equal number of points from $P_s$ in each box. 
\end{definition}
For dimension $s$, 
the largest value of $l$ for which this definition holds is 
called the {\it resolution} of $P_s$ and denoted by $l_s$. 
We have a trivial upper bound $l_s \leq \lfloor m/s \rfloor$. 
As a criterion of uniformity, a high resolution $l_s$ is desirable. 
Thus, we define the \textit{resolution gap} $d_s = \lfloor m/s \rfloor - l_s$ and 
the \textit{sum of resolution gaps} $\Delta = \sum_{s = 1}^m d_s$. 
If $\Delta = 0$, the generator is said to be \textit{fully equidistributed} (\textit{FE}). 
Note that $P_s$ contains the origin $\{ \mathbf{0} \}$ and the output values of a Tausworthe generator for the entire period of $2^m-1$. 
Chen et al.~\cite{MR3173841} implemented FE Tausworthe generators for Markov chain QMC. 

\section{Main result} \label{sec:main}
\subsection{An exhaustive search algorithm using Fibonacci polynomials}

To construct a point set that approximates
CUD sequences in Definition~\ref{def:cud}, 
we search for a pair of polynomials $(p(x), q(x))$ 
whose $t$-values are optimal for $s = 2$ and as small as possible for $s \geq 3$. 
Thus, we refine the algorithm of Tezuka and Fushimi \cite{MR1160278}. 

For dimension $s = 2$, 
there is a connection between the $t$-value of polynomial Korobov lattice point sets (\ref{eqn:Kolobov}) 
and continued fraction expansion of $q(x)/p(x)$.  
Let 
\begin{eqnarray*}
\frac{q(x)}{p(x)} = 
A_0(x) +\cfrac{1}{A_1(x) +\cfrac{1}{A_2(x) +\cfrac{1}{
\ddots \raisebox{-2ex}{$+\cfrac{1}{A_v(x)}$}}}}
=: [A_0(x); A_1(x), A_2(x), \ldots , A_v(x)]
\end{eqnarray*}
be the continued fraction expansion of the rational function $q(x)/p(x)$ 
with a polynomial part $A_0(x) \in \mathbb{F}_2[x]$ and 
partial quotients $A_k(x) \in \mathbb{F}_2[x]$ satisfying $\deg(A_k(x)) \geq 1$ for $1 \leq  k \leq v$. 
\begin{theorem}[\cite{MR1172997,MR1160278}] \label{thm:Tezuka--Fushimi}
Let $p(x) \in \mathbb{F}_2[x]$ with $m = \deg (p(x))$ and $q(x) \in \mathbb{F}_2[x]$ with $\deg (q (x)) < m$. 
Assume $\gcd (p(x),q(x)) = 1$.
Then, the two-dimensional point set
\[ P_2 = \biggl\{  \nu_w \left( \frac{h(x)}{p(x)} (1, q(x)) \right) \Big| \ \deg (h(x)) < m \biggl\} \]
is a $(0, m , 2)$-net (i.e., the $t$-value is zero) if and only 
if the partial quotients in the continued fraction expansion 
$[0; A_1(x), A_2(x), \ldots, A_v(x)]$ of $q(x)/p(x)$ all have degree one, so $v = m$. 
\end{theorem}
The next theorem asserts the existence of $q(x)$ with the above property 
for every irreducible polynomial $p(x)$. 
\begin{theorem}[\cite{MESIROV1987144}]
Let $p(x)$ be an irreducible polynomial with $m = \deg (p(x))$ and $q(x) \in \mathbb{F}_2[x]$ with $\deg q ((x)) < m$. 
For each $p(x)$, there are exactly two polynomials $q(x)$ 
for which the partial quotients of the continued fraction expansion of $q(x)/p(x)$ all have degree one. 
\end{theorem}
In fact, the two polynomials are $q(x)$ and $q^{-1}(x) \mod p(x)$, 
which mean that we generate Tausworthe generators 
in normal order and reverse order, respectively. 
Hence, they yield essentially the same polynomial lattice point set $P_s$. 

\begin{figure}[htbp] 
\begin{center}
\includegraphics[width=12.0cm]{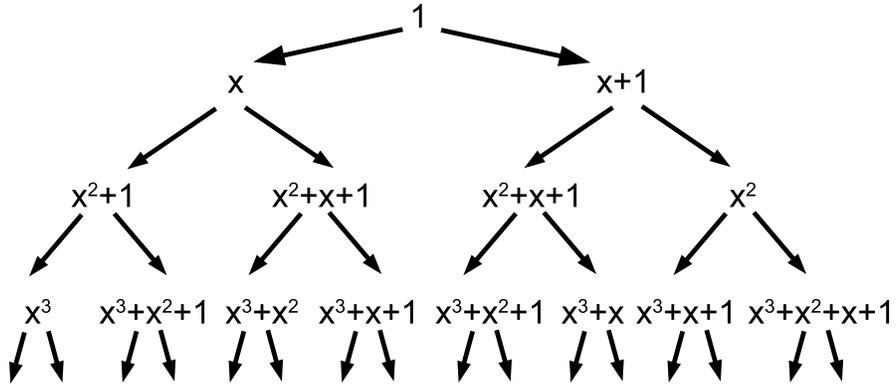}
\end{center}
\caption{A tree of Fibonacci polynomials. \label{fig:Fibonacci_tree}}
\end{figure}

To obtain $(p(x), q(x))$ satisfying the above theorems, 
Tezuka and Fushimi \cite{MR1160278} defined a polynomial analogue of Fibonacci numbers as follows:
 \begin{eqnarray}
F_k(x) & = & A_k(x) F_{k-1}(x) + F_{k-2}(x) \qquad (k \geq 2), \label{eqn:Fibonacci1}\\
F_0(x) & = & 1, \quad F_1(x) = A_1(x), \label{eqn:Fibonacci2}\\
A_k(x)  & = & x \mbox{  or  } x+1 \qquad (k \geq 1) \label{eqn:Fibonacci3}. 
\end{eqnarray}
They called a pair of polynomials $(F_k(x), F_{k-1}(x))$ a pair of ``Fibonacci polynomials" 
because the partial quotients in the continued fraction of $F_{k-1}(x)/F_{k}(x)$ are all of degree one. 
Figure~\ref{fig:Fibonacci_tree} shows the initial part of a tree of Fibonacci polynomials, which was originally illustrated in \cite[Figure~4.5]{rTEZ95a}. 
Note that there are $2^m$ different pairs $(F_m(x), F_{m-1}(x))$ for 
Fibonacci polynomials with degree $m$. 
From them, we choose a suitable pair $(p(x), q(x))$ 
that approximates CUD sequences in Definition~\ref{def:cud}. 

Now we refine the algorithm of Tezuka and Fushimi \cite{MR1160278}. 
Our exhaustive search algorithm proceeds as follows: 
\begin{algorithm}[H] \label{algorithm}
\caption{An exhaustive search algorithm}
\begin{algorithmic}[1]
\State Generate  all the pairs $(F_m(x), F_{m-1}(x))$ using the recurrence relation of Fibonacci polynomials (\ref{eqn:Fibonacci1})--(\ref{eqn:Fibonacci3}). 
\State Check the primitivity of $F_m(x)$.
\State Find $\sigma$ such that $x^{\sigma} \equiv F_{m-1}(x) \mod F_{m}(x)$ and $0 < \sigma < 2^m-1$. Check $\gcd (\sigma, 2^m-1) = 1$ and $\sigma \geq w$.
\State Choose pairs $(F_m(x),F_{m-1}(x))$ whose $t$-value is equal to or smaller than $3$ for $s = 3$. 
\State Let $t^{(s)}$ be a $t$-value for dimension $s$. 
For each $(F_m(x), F_{m-1}(x))$, 
make a vector $(t^{(4)}, t^{(5)}, t^{(6)}, \ldots, t^{(m)})$ of the $t$-values. 
\State Sort pairs $(F_m(x), F_{m-1}(x))$ in lexicographic order based on $(t^{(4)}, t^{(5)}, t^{(6)}, \ldots, t^{(m)})$ starting from dimension $4$. 
\State Choose one of the best (or smallest) pairs $(F_m(x), F_{m-1}(x))$ in Step~6. 
\State Set $(p(x), q(x)) \gets (F_{m}(x), F_{m-1}(x))$. 
\end{algorithmic}
\end{algorithm}

In Step~4, this criterion means that the $t$-value is sufficiently small for $s = 3$; see Remark~\ref{remark:t-value} for details. 
In Steps~4 and 5, we calculate the $t$-values by using Gaussian elimination \cite{Pirsic2001827} 
instead of solving Diophantine equations in \cite[Theorem~1]{MR1160278}. 
\begin{remark} \label{remark:trinomial}
In the original paper \cite{MR1160278}, 
before Step~2, 
Tezuka and Fushimi checked the condition 
\[ F_{m-1}(x)^m + F_{m-1}(x)^n + 1 = 0 \mod{F_m(x)}, \]
where $0 < n < m$, to obtain fast Tausworthe generators 
using trinomial generalized feedback shift register generators. 
They also restricted the calculation of  the $t$-values to only $3 \leq s \leq 6$. 
A reason for these conditions might be the difficulty of checking from Steps~2--5 
on computers around 1990. 
As a result, in the range $3 \leq m \leq 32$, 
there exist pairs $(F_m(x), F_{m-1}(x))$ only for 
$ m = 3, 5, 7, 15, 17, 18, 20, 22, 23, 25, 28$, and $31$; 
otherwise, there exists no pair. 
In the related paper \cite{Tezuka:1992:FGL:167293.167397}, 
the authors found pairs $(F_m(x), F_{m-1}(x))$ 
for all $3 \leq m \leq 21$ under a pentanomial condition. 
Currently, it is not difficult to remove these conditions when we conduct an exhaustive search 
on modern computers. 
In Remark~\ref{remark:generation}, we note a reasonably fast generation method  instead of the direct use of Definition~\ref{def:Tausworthe}. 
\end{remark}

\begin{remark} \label{remark:t-value}
In Step~4, we observed that the smallest $t$-values are $2$ or $3$ for $10 \leq s \leq 32$ by exhaustive search. 
More precisely, there exist pairs $(p(x), q(x))$ with $t$-value two only for $10 \leq s \leq 14$ and $s = 16$ and $17$, and the number of them are quite few, compared with the number of pairs with $t$-value three. 
For example, in the case where $s = 17$, there exist four pairs with $t$-value two but 464 
pairs with $t$-value three. Thus, to find a pair $(p(x), q(x))$ with smaller $t$-value even for $s \geq 4$, we adopted this criterion. 
\end{remark}


\subsection{Specific parameters}

Table~\ref{table:parameter} lists specific parameters for $w = 32, 64$ and $10 \leq m \leq 32$. 
In Table~\ref{table:parameter}, each first and second row shows 
the coefficients of $p(x)$ and $q(x)$ 
respectively; for example, $1 \ 1 \ 0 \ 1$ means $1 + x + x^3$. 
We also note the step size $\sigma$ corresponding to $q(x)$. 
For $m = 21$ and $28$, we obtained the pairs of polynomials $(p(x), q(x))$ 
with somewhat large defects $\Delta = 6$ and $4$, respectively, 
so we replaced them by the second-best pairs. 
Table~\ref{table:t-value} summarizes the 
$t$-values and sum of resolution gaps $\Delta$ for our new Tausworthe generators (labeled ``New")
and the existing Tausworthe generators developed by Chen et al.~\cite{MR3173841} (labeled ``Chen") in the range of $2 \leq s \leq 20$. 
For $2 \leq s \leq 5$, our new generators have the $t$-values equal to or smaller than the existing generators (except for $m = 32$). 
It is known that QMC are successful in high-dimensional problems, 
particularly in the case in which problems are dominated by the first few variables, 
so we focus on the optimization of leading dimensions. 
Conversely, from the viewpoint of the FE property, our generators are not FE. 
We can also optimize both the $t$-values and FE property, 
but the $t$-values slightly increase. 
Thus, we prioritized the $t$-values over the FE property for simplicity. 
The code in C is available at \url{https://github.com/sharase/cud}.

\begin{table}[htbp] 
\caption{Specific parameters of pairs of polynomials $(p(x), q(x))$ and step sizes $\sigma$. } \label{table:parameter}
{\scriptsize
  \begin{tabular}{|l|l|} \hline


$m = 10$ & $1 \ 0 \ 0 \ 0 \ 0 \ 0 \ 1 \ 1 \ 0 \ 1 \ 1$ \\ 
{} & $0 \ 1 \ 0 \ 1 \ 1 \ 1 \ 0 \ 1 \ 0 \ 1$ \quad ($\sigma = 70$) \\ \hline  

$m = 11$ & $1 \ 1 \ 0 \ 0 \ 1 \ 0 \ 0 \ 1 \ 1 \ 0 \ 1 \ 1$ \\ 
{} & $0 \ 1 \ 0 \ 0 \ 0 \ 0 \ 1 \ 1 \ 1 \ 0 \ 1$ \quad ($\sigma = 179$) \\ \hline  

$m = 12$ & $1 \ 1 \ 1 \ 1 \ 1 \ 0 \ 0 \ 1 \ 0 \ 0 \ 1 \ 1 \ 1$ \\ 
{} & $0 \ 0 \ 1 \ 0 \ 0 \ 1 \ 1 \ 1 \ 1 \ 0 \ 1 \ 1$ \quad ($\sigma = 146$) \\ \hline  

$m = 13$ & $1 \ 1 \ 1 \ 0 \ 1 \ 0 \ 0 \ 0 \ 1 \ 0 \ 1 \ 1 \ 1 \ 1$ \\ 
{} & $1 \ 0 \ 1 \ 0 \ 1 \ 1 \ 1 \ 1 \ 1 \ 0 \ 0 \ 1 \ 1$ \quad ($\sigma = 139$) \\ \hline  

$m = 14$ & $1 \ 0 \ 1 \ 0 \ 1 \ 1 \ 0 \ 1 \ 1 \ 1 \ 1 \ 0 \ 1 \ 1 \ 1$ \\ 
{} & $1 \ 0 \ 1 \ 1 \ 1 \ 1 \ 0 \ 1 \ 0 \ 0 \ 1 \ 0 \ 1 \ 1$ \quad ($\sigma = 5192$) \\ \hline  

$m = 15$ & $1 \ 1 \ 0 \ 1 \ 1 \ 0 \ 0 \ 1 \ 1 \ 1 \ 0 \ 1 \ 0 \ 1 \ 1 \ 1$ \\ 
{} & $0 \ 0 \ 1 \ 1 \ 0 \ 1 \ 1 \ 1 \ 0 \ 0 \ 0 \ 0 \ 0 \ 1 \ 1$ \quad ($\sigma = 1028$) \\ \hline  

$m = 16$ & $1 \ 1 \ 0 \ 1 \ 0 \ 1 \ 1 \ 1 \ 1 \ 1 \ 0 \ 0 \ 1 \ 0 \ 0 \ 1 \ 1$ \\ 
{} & $1 \ 0 \ 0 \ 1 \ 1 \ 1 \ 0 \ 1 \ 0 \ 0 \ 1 \ 1 \ 0 \ 1 \ 1 \ 1$ \quad ($\sigma = 12749$) \\ \hline  

$m = 17$ & $1 \ 0 \ 1 \ 1 \ 1 \ 0 \ 0 \ 0 \ 0 \ 1 \ 0 \ 1 \ 1 \ 0 \ 0 \ 0 \ 1 \ 1$ \\ 
{} & $1 \ 1 \ 1 \ 1 \ 0 \ 1 \ 0 \ 1 \ 1 \ 1 \ 0 \ 1 \ 1 \ 1 \ 1 \ 0 \ 1$ \quad ($\sigma = 20984$) \\ \hline  

$m = 18$ & $1 \ 1 \ 0 \ 1 \ 0 \ 1 \ 1 \ 0 \ 1 \ 0 \ 1 \ 0 \ 0 \ 0 \ 1 \ 1 \ 0 \ 1 \ 1$ \\ 
{} & $1 \ 1 \ 1 \ 0 \ 0 \ 1 \ 1 \ 1 \ 0 \ 0 \ 0 \ 0 \ 0 \ 1 \ 1 \ 1 \ 0 \ 1$ \quad ($\sigma = 72349$) \\ \hline  

$m = 19$ & $1 \ 0 \ 1 \ 1 \ 0 \ 1 \ 1 \ 1 \ 1 \ 0 \ 0 \ 0 \ 1 \ 1 \ 0 \ 0 \ 1 \ 0 \ 0 \ 1$ \\ 
{} & $0 \ 0 \ 0 \ 0 \ 1 \ 1 \ 1 \ 1 \ 0 \ 0 \ 0 \ 0 \ 1 \ 1 \ 1 \ 0 \ 1 \ 0 \ 1$ \quad ($\sigma = 92609$) \\ \hline  

$m = 20$ & $1 \ 1 \ 1 \ 0 \ 1 \ 0 \ 1 \ 0 \ 1 \ 1 \ 1 \ 0 \ 0 \ 1 \ 1 \ 1 \ 0 \ 0 \ 1 \ 0 \ 1$ \\ 
{} & $0 \ 1 \ 0 \ 0 \ 0 \ 1 \ 1 \ 1 \ 1 \ 0 \ 0 \ 1 \ 1 \ 1 \ 0 \ 0 \ 1 \ 0 \ 0 \ 1$ \quad ($\sigma = 226826$) \\ \hline  

$m = 21$ & $1 \ 1 \ 1 \ 1 \ 1 \ 1 \ 0 \ 1 \ 1 \ 1 \ 0 \ 0 \ 1 \ 0 \ 1 \ 0 \ 1 \ 1 \ 1 \ 0 \ 0 \ 1$ \\  
{} & $0 \ 1 \ 0 \ 1 \ 1 \ 1 \ 0 \ 0 \ 1 \ 1 \ 0 \ 0 \ 1 \ 1 \ 0 \ 1 \ 0 \ 0 \ 1 \ 0 \ 1$ \quad ($\sigma = 1127911$) \\ \hline  

$m = 22$ & $1 \ 1 \ 0 \ 0 \ 1 \ 0 \ 0 \ 0 \ 1 \ 1 \ 0 \ 0 \ 1 \ 0 \ 1 \ 0 \ 0 \ 0 \ 1 \ 1 \ 0 \ 1 \ 1$ \\ 
{} & $0 \ 0 \ 1 \ 1 \ 0 \ 1 \ 0 \ 0 \ 0 \ 0 \ 1 \ 0 \ 0 \ 1 \ 0 \ 0 \ 1 \ 1 \ 0 \ 1 \ 1 \ 1$ \quad ($\sigma = 629680$) \\ \hline  

$m = 23$ & $1 \ 1 \ 1 \ 0 \ 0 \ 1 \ 1 \ 0 \ 0 \ 1 \ 0 \ 1 \ 0 \ 1 \ 1 \ 0 \ 0 \ 1 \ 1 \ 1 \ 0 \ 0 \ 0 \ 1$ \\ 
{} & $1 \ 0 \ 1 \ 0 \ 0 \ 1 \ 0 \ 0 \ 1 \ 1 \ 0 \ 0 \ 1 \ 0 \ 1 \ 1 \ 1 \ 1 \ 0 \ 0 \ 0 \ 1 \ 1$ \quad ($\sigma = 1796311$) \\ \hline  

$m = 24$ & $1 \ 1 \ 1 \ 1 \ 0 \ 0 \ 0 \ 1 \ 1 \ 0 \ 1 \ 0 \ 1 \ 1 \ 0 \ 0 \ 0 \ 1 \ 0 \ 1 \ 1 \ 1 \ 1 \ 0 \ 1$ \\ 
{} & $1 \ 1 \ 0 \ 0 \ 0 \ 0 \ 1 \ 1 \ 1 \ 1 \ 1 \ 1 \ 0 \ 0 \ 1 \ 0 \ 1 \ 0 \ 1 \ 0 \ 1 \ 1 \ 1 \ 1$ \quad ($\sigma = 7017398$) \\ \hline  

$m = 25$ & $1 \ 1 \ 1 \ 0 \ 1 \ 0 \ 1 \ 1 \ 0 \ 0 \ 1 \ 1 \ 0 \ 1 \ 1 \ 0 \ 0 \ 1 \ 0 \ 1 \ 1 \ 0 \ 1 \ 1 \ 1 \ 1$ \\ 
{} & $0 \ 1 \ 0 \ 1 \ 0 \ 0 \ 1 \ 0 \ 0 \ 0 \ 1 \ 0 \ 1 \ 0 \ 0 \ 1 \ 1 \ 1 \ 0 \ 1 \ 1 \ 0 \ 0 \ 1 \ 1$ \quad ($\sigma = 2947446$) \\ \hline  

$m = 26$ & $1 \ 1 \ 1 \ 0 \ 1 \ 0 \ 1 \ 1 \ 0 \ 1 \ 0 \ 1 \ 1 \ 0 \ 1 \ 1 \ 1 \ 0 \ 0 \ 0 \ 0 \ 0 \ 1 \ 1 \ 1 \ 1 \ 1$ \\ 
{} & $1 \ 1 \ 0 \ 1 \ 1 \ 1 \ 0 \ 1 \ 0 \ 0 \ 0 \ 0 \ 1 \ 0 \ 1 \ 1 \ 0 \ 1 \ 0 \ 0 \ 0 \ 0 \ 0 \ 0 \ 1 \ 1$ \quad ($\sigma = 19101221$) \\ \hline  

$m = 27$ & $1 \ 1 \ 0 \ 0 \ 0 \ 1 \ 0 \ 0 \ 1 \ 0 \ 0 \ 0 \ 1 \ 0 \ 1 \ 0 \ 0 \ 0 \ 1 \ 1 \ 0 \ 1 \ 1 \ 1 \ 0 \ 1 \ 0 \ 1$ \\ 
{} & $0 \ 1 \ 0 \ 1 \ 0 \ 0 \ 0 \ 1 \ 1 \ 1 \ 1 \ 1 \ 1 \ 0 \ 1 \ 0 \ 1 \ 0 \ 0 \ 1 \ 0 \ 1 \ 0 \ 1 \ 1 \ 1 \ 1$ \quad ($\sigma = 4397933$) \\ \hline  

$m = 28$ & $1 \ 0 \ 0 \ 0 \ 1 \ 0 \ 1 \ 1 \ 0 \ 0 \ 0 \ 1 \ 1 \ 0 \ 1 \ 0 \ 1 \ 0 \ 0 \ 1 \ 1 \ 0 \ 0 \ 1 \ 0 \ 1 \ 1 \ 1 \ 1$ \\ 
{} & $0 \ 0 \ 0 \ 1 \ 1 \ 0 \ 1 \ 0 \ 0 \ 1 \ 1 \ 0 \ 0 \ 0 \ 1 \ 1 \ 1 \ 1 \ 0 \ 0 \ 0 \ 1 \ 0 \ 1 \ 0 \ 0 \ 1 \ 1$ \quad ($\sigma = 167713336$) \\ \hline  

$m = 29$ & $1 \ 0 \ 1 \ 0 \ 0 \ 0 \ 0 \ 0 \ 0 \ 1 \ 0 \ 1 \ 0 \ 1 \ 0 \ 1 \ 1 \ 0 \ 1 \ 1 \ 1 \ 0 \ 0 \ 1 \ 1 \ 0 \ 1 \ 0 \ 1 \ 1$ \\ 
{} & $1 \ 1 \ 1 \ 1 \ 1 \ 0 \ 1 \ 0 \ 0 \ 1 \ 1 \ 1 \ 0 \ 0 \ 0 \ 0 \ 1 \ 0 \ 1 \ 1 \ 1 \ 1 \ 0 \ 1 \ 0 \ 1 \ 1 \ 0 \ 1$ \quad ($\sigma = 83189117$) \\ \hline  

$m = 30$ & $1 \ 0 \ 0 \ 0 \ 0 \ 1 \ 0 \ 1 \ 1 \ 0 \ 0 \ 0 \ 1 \ 0 \ 1 \ 0 \ 0 \ 1 \ 1 \ 1 \ 1 \ 1 \ 0 \ 0 \ 0 \ 0 \ 0 \ 1 \ 0 \ 0 \ 1$ \\ 
{} & $0 \ 1 \ 0 \ 1 \ 1 \ 1 \ 1 \ 0 \ 1 \ 0 \ 0 \ 0 \ 0 \ 0 \ 0 \ 0 \ 1 \ 1 \ 0 \ 0 \ 0 \ 1 \ 1 \ 1 \ 1 \ 0 \ 1 \ 1 \ 0 \ 1$ \quad ($\sigma = 315800840$) \\ \hline  

$m = 31$ & $1 \ 0 \ 1 \ 1 \ 1 \ 0 \ 1 \ 1 \ 1 \ 0 \ 0 \ 0 \ 0 \ 1 \ 0 \ 0 \ 0 \ 0 \ 1 \ 1 \ 1 \ 0 \ 1 \ 1 \ 1 \ 1 \ 0 \ 1 \ 1 \ 0 \ 1 \ 1$ \\ 
{} & $0 \ 0 \ 0 \ 0 \ 1 \ 1 \ 1 \ 1 \ 0 \ 1 \ 0 \ 0 \ 0 \ 1 \ 1 \ 0 \ 1 \ 1 \ 1 \ 1 \ 1 \ 1 \ 1 \ 0 \ 0 \ 1 \ 1 \ 0 \ 1 \ 0 \ 1$ \quad ($\sigma = 36109125$) \\ \hline  

$m = 32$ & $1 \ 0 \ 0 \ 0 \ 1 \ 0 \ 1 \ 0 \ 1 \ 1 \ 0 \ 1 \ 1 \ 1 \ 1 \ 1 \ 1 \ 1 \ 0 \ 0 \ 0 \ 0 \ 0 \ 1 \ 0 \ 1 \ 0 \ 0 \ 0 \ 1 \ 1 \ 0 \ 1$ \\ 
{} & $0 \ 1 \ 0 \ 0 \ 0 \ 0 \ 1 \ 1 \ 1 \ 0 \ 1 \ 1 \ 1 \ 0 \ 1 \ 1 \ 0 \ 1 \ 0 \ 1 \ 0 \ 1 \ 0 \ 1 \ 0 \ 1 \ 1 \ 1 \ 1 \ 1 \ 1 \ 1$ \quad ($\sigma = 686019401$) \\ \hline  
 \end{tabular}
}
\end{table}

\begin{table}[htbp]
\caption{Comparison of the $t$-values and $\Delta$ for our new Tausworthe generators and the existing Tausworthe generators developed by Chen et al.~\cite{MR3173841}.} \label{table:t-value}
{\scriptsize
  \begin{tabular}{|c|c|rrrrrrrrrrrrrrrrrrr|r|} \hline
$m$ & dim. $s$ & 2 & 3 & 4 & 5 & 6 & 7 & 8 & 9 & 10 & 11 & 12 & 13 & 14 & 15 & 16 & 17 & 18 & 19 & 20 & $\Delta$ \\ \hline
$10$ & New & 0 & 3 & 3 & 4 & 5 & 5 & 6 & 6 & 6 & 6 & 6 & 6 & 6 & 6 & 6 & 6 & 6 & 6 & 7 & 2 \\
{} & Chen & 2 & 5 & 5 & 5 & 6 & 6 & 6 & 7 & 7 & 7 & 7 & 7 & 7 & 7 & 7 & 7 & 7 & 7 & 7 & 0 \\ \hline
$11$ & New & 0 & 3 & 3 & 5 & 6 & 6 & 6 & 6 & 7 & 7 & 7 & 7 & 7 & 7 & 7 & 7 & 7 & 7 & 7 & 1 \\
{} & Chen & 2 & 5 & 5 & 6 & 6 & 6 & 7 & 7 & 7 & 7 & 7 & 7 & 7 & 7 & 8 & 8 & 8 & 8 & 8 & 0 \\ \hline
$12$ & New & 0 & 3 & 4 & 5 & 6 & 6 & 6 & 6 & 6 & 6 & 6 & 8 & 8 & 8 & 8 & 8 & 8 & 8 & 8 & 2 \\
{} & Chen & 2 & 3 & 5 & 5 & 7 & 7 & 7 & 7 & 7 & 7 & 7 & 7 & 8 & 8 & 8 & 8 & 8 & 8 & 8 & 0 \\ \hline
$13$ & New & 0 & 2 & 3 & 5 & 6 & 6 & 7 & 7 & 7 & 8 & 8 & 8 & 8 & 8 & 9 & 9 & 9 & 9 & 9 & 0 \\
{} & Chen & 1 & 5 & 5 & 5 & 6 & 8 & 9 & 9 & 9 & 9 & 9 & 9 & 9 & 9 & 9 & 9 & 9 & 9 & 9 & 0 \\ \hline
$14$ & New & 0 & 3 & 4 & 5 & 7 & 7 & 7 & 7 & 8 & 9 & 9 & 9 & 9 & 9 & 9 & 9 & 9 & 9 & 9 & 1 \\
{} & Chen & 1 & 6 & 7 & 7 & 7 & 7 & 8 & 9 & 9 & 9 & 9 & 9 & 9 & 9 & 9 & 9 & 10 & 10 & 10 & 0 \\ \hline
$15$ & New & 0 & 3 & 4 & 6 & 7 & 8 & 8 & 9 & 9 & 9 & 9 & 10 & 10 & 10 & 10 & 10 & 10 & 10 & 10 & 1 \\
{} & Chen & 2 & 4 & 5 & 7 & 7 & 7 & 8 & 8 & 9 & 9 & 9 & 9 & 9 & 9 & 9 & 9 & 9 & 10 & 10 & 0 \\ \hline
$16$ & New & 0 & 3 & 4 & 7 & 7 & 8 & 10 & 10 & 10 & 11 & 11 & 11 & 11 & 11 & 11 & 11 & 11 & 11 & 11 & 1 \\
{} & Chen & 3 & 4 & 5 & 8 & 8 & 8 & 8 & 8 & 10 & 10 & 10 & 10 & 10 & 10 & 10 & 10 & 10 & 10 & 12 & 0 \\ \hline
$17$ & New & 0 & 3 & 4 & 7 & 7 & 7 & 8 & 10 & 10 & 10 & 10 & 11 & 11 & 11 & 11 & 11 & 12 & 12 & 12 & 1 \\
{} & Chen & 2 & 5 & 6 & 10 & 10 & 10 & 10 & 10 & 10 & 10 & 10 & 10 & 10 & 10 & 10 & 11 & 11 & 11 & 11 & 0 \\ \hline
$18$ & New & 0 & 3 & 5 & 6 & 7 & 9 & 9 & 9 & 10 & 10 & 10 & 10 & 11 & 11 & 11 & 12 & 12 & 13 & 13 & 2 \\
{} & Chen & 3 & 4 & 5 & 7 & 8 & 9 & 9 & 12 & 12 & 12 & 12 & 12 & 12 & 12 & 12 & 12 & 12 & 12 & 12 & 0 \\ \hline
$19$ & New & 0 & 3 & 5 & 6 & 7 & 12 & 12 & 12 & 12 & 12 & 12 & 12 & 13 & 13 & 13 & 13 & 13 & 13 & 13 & 1 \\
{} & Chen & 2 & 4 & 8 & 8 & 8 & 9 & 9 & 9 & 11 & 12 & 12 & 12 & 12 & 12 & 12 & 12 & 12 & 12 & 12 & 0 \\ \hline
$20$ & New & 0 & 3 & 5 & 7 & 7 & 10 & 10 & 11 & 11 & 12 & 12 & 13 & 13 & 13 & 13 & 13 & 13 & 13 & 13 & 2 \\
{} & Chen & 3 & 4 & 8 & 8 & 8 & 13 & 13 & 13 & 13 & 13 & 13 & 13 & 13 & 14 & 14 & 14 & 14 & 14 & 14 & 0 \\ \hline
$21$ & New & 0 & 3 & 5 & 8 & 8 & 9 & 10 & 10 & 10 & 13 & 13 & 13 & 13 & 13 & 13 & 13 & 13 & 14 & 14 & 1 \\
{} & Chen & 3 & 6 & 8 & 8 & 8 & 11 & 11 & 11 & 12 & 12 & 12 & 12 & 12 & 12 & 12 & 12 & 13 & 13 & 15 & 0 \\ \hline
$22$ & New & 0 & 3 & 5 & 7 & 10 & 10 & 12 & 12 & 12 & 12 & 13 & 13 & 13 & 13 & 15 & 15 & 15 & 15 & 15 & 1 \\
{} & Chen & 7 & 7 & 7 & 8 & 8 & 14 & 14 & 14 & 14 & 14 & 14 & 14 & 14 & 14 & 14 & 14 & 14 & 14 & 15 & 0 \\ \hline
$23$ & New & 0 & 3 & 5 & 9 & 9 & 11 & 12 & 13 & 13 & 13 & 13 & 13 & 13 & 13 & 15 & 15 & 15 & 15 & 15 & 1 \\
{} & Chen & 5 & 5 & 9 & 9 & 9 & 9 & 11 & 15 & 15 & 15 & 15 & 15 & 15 & 15 & 15 & 15 & 15 & 15 & 15 & 0 \\ \hline
$24$ & New & 0 & 3 & 6 & 8 & 10 & 11 & 12 & 13 & 14 & 14 & 14 & 14 & 15 & 17 & 17 & 17 & 17 & 17 & 17 & 3 \\
{} & Chen & 5 & 5 & 8 & 8 & 11 & 11 & 11 & 12 & 14 & 14 & 14 & 14 & 14 & 14 & 14 & 15 & 15 & 16 & 16 & 0 \\ \hline
$25$ & New & 0 & 3 & 6 & 7 & 12 & 12 & 12 & 13 & 13 & 13 & 14 & 14 & 16 & 16 & 16 & 18 & 18 & 18 & 18 & 3 \\
{} & Chen & 4 & 6 & 8 & 8 & 9 & 10 & 11 & 12 & 12 & 12 & 14 & 16 & 16 & 16 & 16 & 16 & 16 & 16 & 16 & 0 \\ \hline
$26$ & New & 0 & 3 & 6 & 8 & 12 & 12 & 12 & 13 & 13 & 13 & 14 & 14 & 15 & 15 & 15 & 16 & 16 & 16 & 18 & 2 \\
{} & Chen & 6 & 7 & 7 & 9 & 11 & 11 & 12 & 13 & 13 & 14 & 15 & 15 & 16 & 16 & 16 & 16 & 17 & 17 & 17 & 0 \\ \hline
$27$ & New & 0 & 3 & 7 & 7 & 11 & 12 & 13 & 13 & 13 & 14 & 14 & 14 & 16 & 16 & 16 & 16 & 16 & 16 & 16 & 3 \\
{} & Chen & 3 & 6 & 8 & 11 & 12 & 12 & 14 & 14 & 14 & 15 & 15 & 15 & 15 & 15 & 16 & 16 & 16 & 17 & 17 & 0 \\ \hline
$28$ & New & 0 & 3 & 7 & 9 & 9 & 13 & 13 & 13 & 13 & 14 & 15 & 17 & 17 & 17 & 17 & 17 & 17 & 17 & 17 & 2 \\
{} & Chen & 4 & 5 & 13 & 13 & 13 & 13 & 13 & 14 & 15 & 15 & 15 & 16 & 16 & 16 & 17 & 17 & 17 & 18 & 18 & 0 \\ \hline
$29$ & New & 0 & 3 & 6 & 9 & 11 & 13 & 14 & 14 & 14 & 20 & 20 & 20 & 20 & 20 & 20 & 20 & 20 & 20 & 20 & 1 \\
{} & Chen & 5 & 5 & 12 & 12 & 12 & 12 & 14 & 14 & 15 & 17 & 17 & 17 & 17 & 17 & 17 & 17 & 17 & 17 & 18 & 0 \\ \hline
$30$ & New & 0 & 3 & 7 & 9 & 12 & 13 & 14 & 14 & 16 & 16 & 16 & 17 & 17 & 17 & 17 & 17 & 17 & 18 & 19 & 1 \\
{} & Chen & 2 & 7 & 7 & 10 & 13 & 13 & 13 & 14 & 17 & 17 & 17 & 17 & 17 & 17 & 18 & 18 & 18 & 18 & 19 & 0 \\ \hline
$31$ & New & 0 & 3 & 7 & 9 & 12 & 12 & 15 & 15 & 15 & 16 & 18 & 19 & 19 & 19 & 19 & 19 & 19 & 19 & 20 & 1 \\
{} & Chen & 2 & 5 & 9 & 10 & 13 & 13 & 15 & 15 & 15 & 15 & 17 & 18 & 18 & 18 & 18 & 18 & 19 & 19 & 19 & 0 \\ \hline
$32$ & New & 0 & 3 & 7 & 10 & 13 & 14 & 14 & 15 & 15 & 17 & 17 & 17 & 18 & 18 & 20 & 20 & 20 & 20 & 20 & 4 \\
{} & Chen & 5 & 5 & 9 & 9 & 13 & 13 & 15 & 15 & 15 & 15 & 16 & 16 & 17 & 18 & 18 & 18 & 19 & 19 & 20 & 0 \\ \hline
  \end{tabular}}
\end{table}

\begin{remark} \label{remark:generation}
We note a reasonably fast generation method for Tausworthe generators. 
Let $\mathbf{x}_i = (a_{i \sigma}, a_{i \sigma +1}, \ldots, a_{i \sigma +m-1}, a_{i \sigma + m}, \ldots, a_{i \sigma + w-1})^\mathsf{T}$ be a $w$-bit state vector at step $i$ for $m \leq w$.  We can define a state transition 
$\mathbf{x}_{i + 1} = \mathbf{B} \mathbf{x}_{i}$, 
where $\mathbf{B} := \begin{pmatrix}
 \mathbf{b}_0 & \ldots & \mathbf{b}_{m-1} & \mathbf{0} & \ldots & \mathbf{0}
\end{pmatrix}$ is a $w \times w$ state transition matrix consisting of $w$-bit column vectors
$\mathbf{b}_0, \ldots,  \mathbf{b}_{m-1}$ and $w-m$ $w$-bit zero column vectors $\mathbf{0}$. 
Then, we have the recurrence relation 
 $\mathbf{x}_{i+1} = a_{i \sigma} \mathbf{b}_0\oplus a_{i \sigma +1} \mathbf{b}_1 \oplus \cdots \oplus a_{i \sigma +m-1} \mathbf{b}_{m-1}$,
which can be calculated by adding column vectors $\mathbf{b}_j$ if $a_{i \sigma +j} = 1$ holds for $j = 0, \ldots, m-1$, 
where the symbol $\oplus$ denotes the bitwise exclusive-or operation. 
Using this method, we can generate $\{ u_i \}_{i = 0}^{\infty}$ in (\ref{eqn:Tausworthe}) with reasonable speed. 
See \cite[\S3 and 5.1]{LP2009} for the construction of $B$. 
\end{remark}

\section{Numerical examples} \label{sec:example}

In this section, we provide numerical examples 
to confirm the performance of Markov chain QMC. 

\subsection{Two-dimensional Gaussian Gibbs sampling} \label{subsec:gaussian}

Our first example is 
a systematic Gibbs sampler to generate the two-dimensional Gaussian distribution 
\[ \mathbf{X}= \begin{pmatrix} 
X_1 \\
X_2
\end{pmatrix} \sim 
\mathcal{N} \left( 
\begin{pmatrix} 
0 \\
0
\end{pmatrix}, 
\begin{pmatrix} 
1 & \rho \\
\rho & 1
\end{pmatrix}
\right) 
\]
for correlation $\rho \in (-1, 1)$. This can be implemented as 
\begin{eqnarray*}
X_{i,1} & \gets & \rho X_{i-1,2} + \sqrt{1-\rho^2} \Phi^{-1}(u_{2i-2}), \label{eqn:sibbs1} \\
X_{i,2} & \gets & \rho X_{i,1} + \sqrt{1-\rho^2} \Phi^{-1}(u_{2i-1}), \label{eqn:gibbs}
\end{eqnarray*}
where $\Phi$ is the cumulative distribution function 
for the standard normal distribution. 
For the output values (\ref{eqn:Tausworthe sequence}) generated from Tausworthe generators, 
we define two-dimensional non-overlapping points starting from the origin:
\begin{eqnarray} \label{eqn:gibbs_output}
 (0, 0), (u_0, u_1), (u_2, u_3), \ldots, (u_{N-2}, u_{0}), (u_1, u_2), \ldots, (u_{N-3}, u_{N-2}),
 \end{eqnarray}
where $N = 2^m$. 
We apply digital shifts, that is, 
we add $(z_1, z_2)$ to each point in (\ref{eqn:gibbs_output}) 
using bitwise exclusive-or $\oplus$, where $z_1$ and $z_2$ are IID samples from U(0, 1). 

\begin{figure}[htbp] \label{fig:gaussian}
\includegraphics[width=8.0cm]{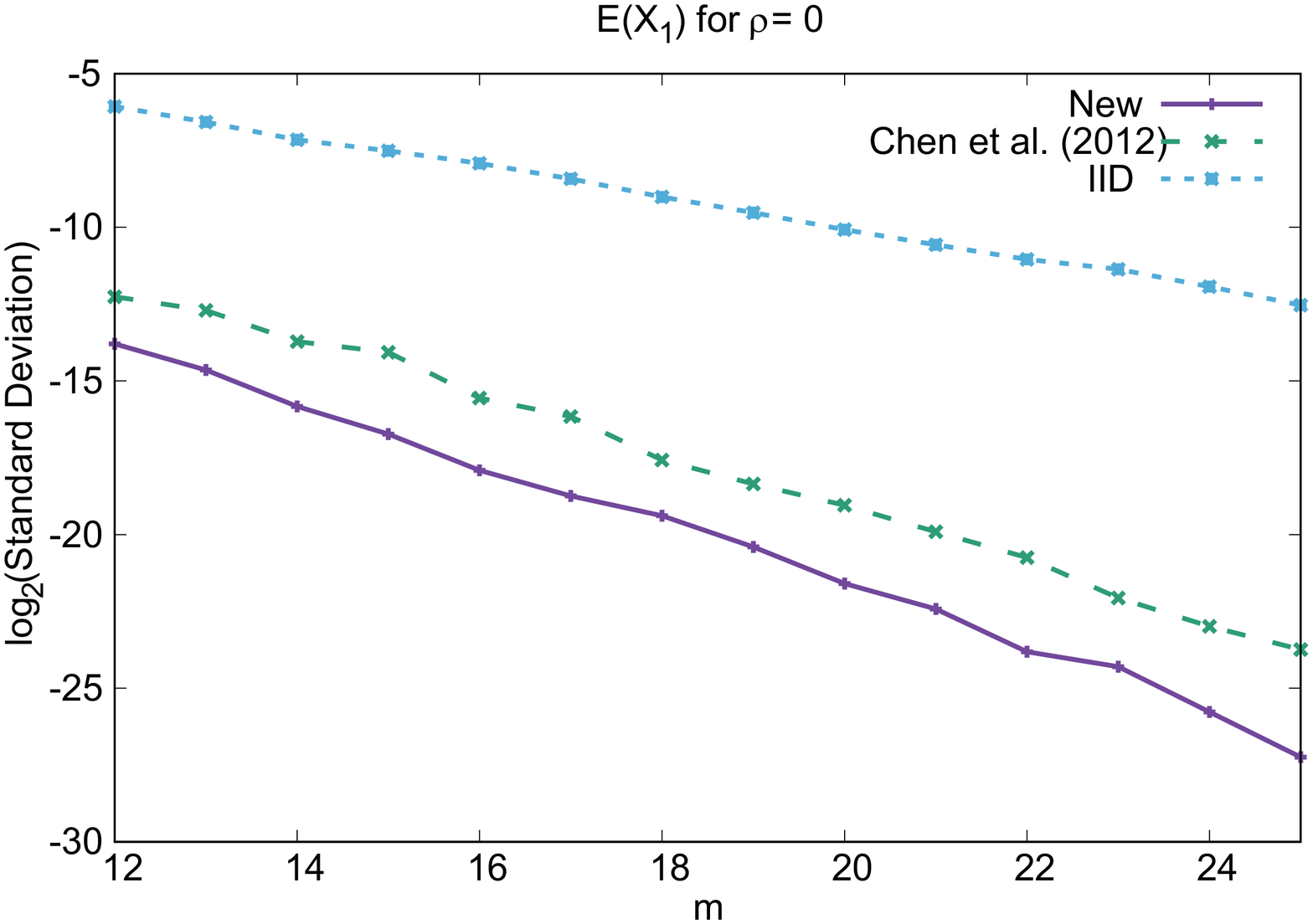}
\includegraphics[width=8.0cm]{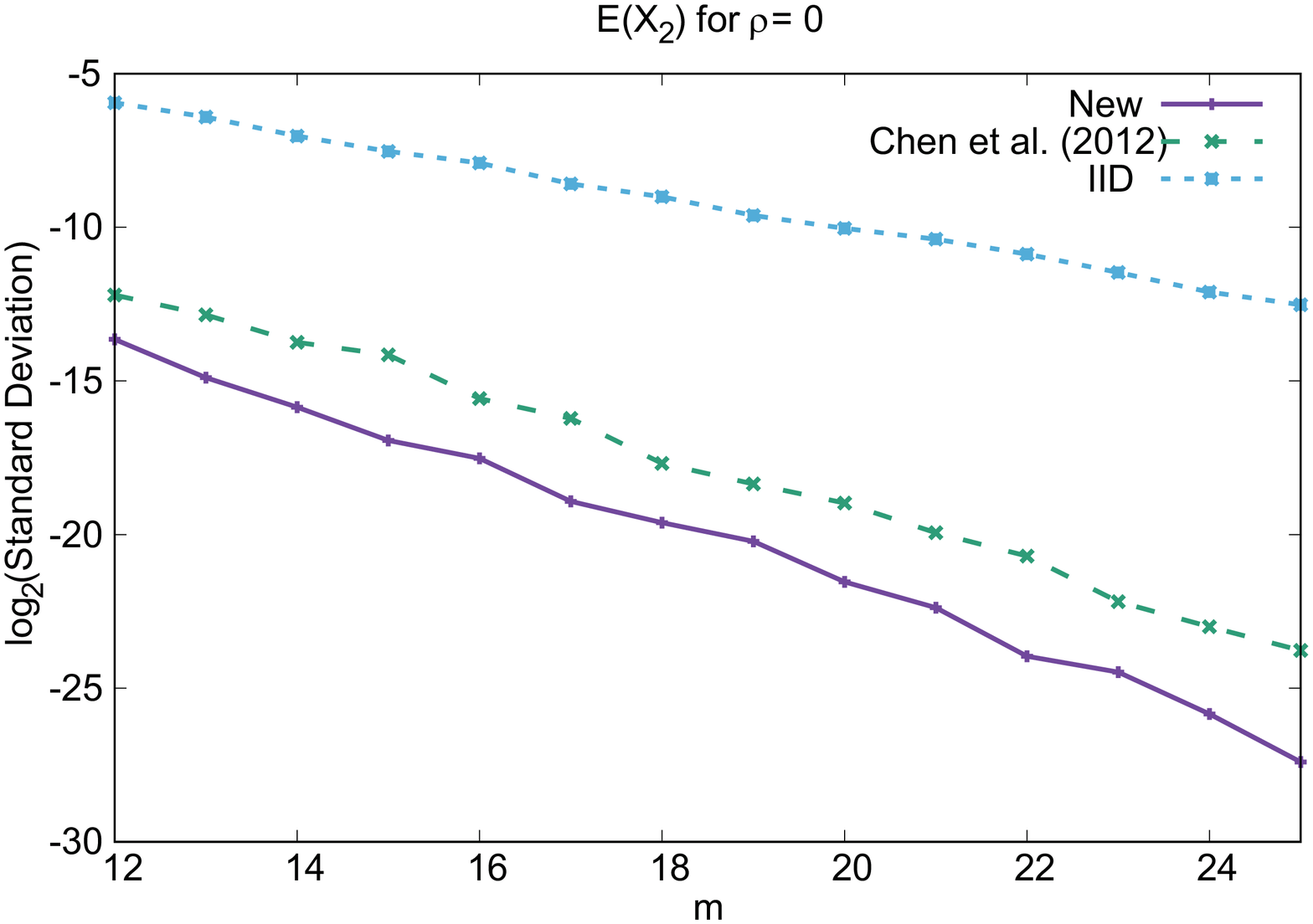}\\
\includegraphics[width=8.0cm]{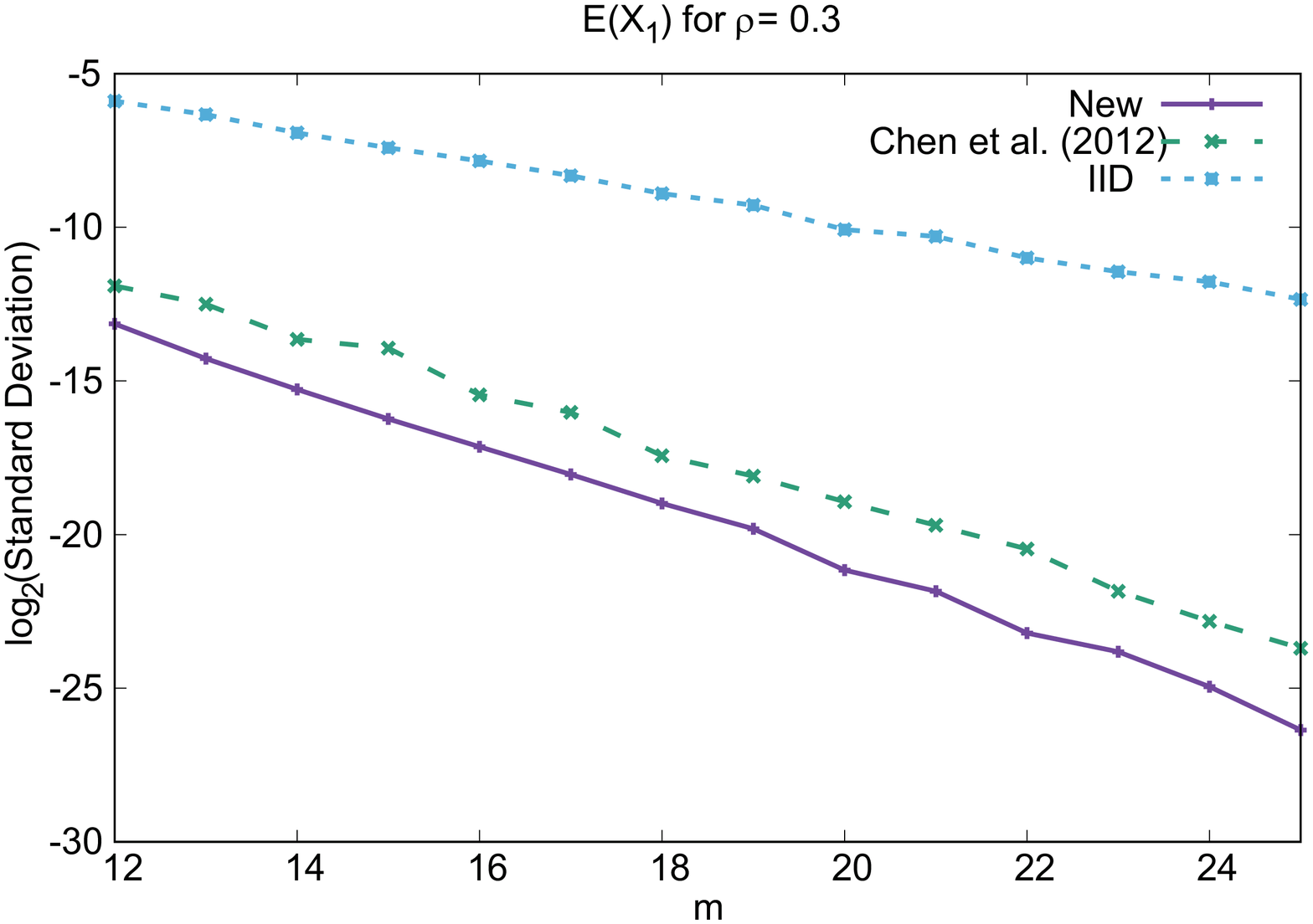}
\includegraphics[width=8.0cm]{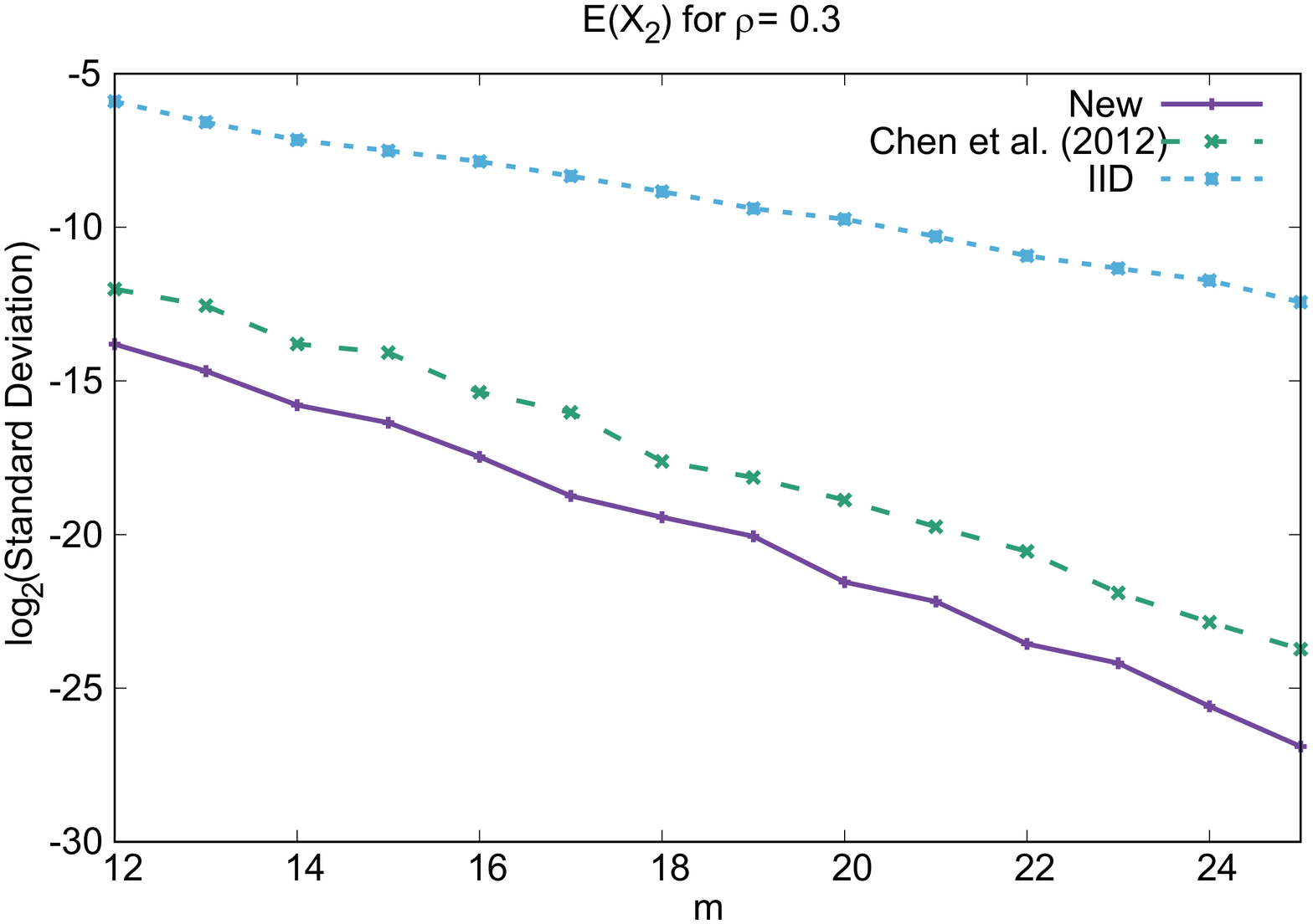}\\
\includegraphics[width=8.0cm]{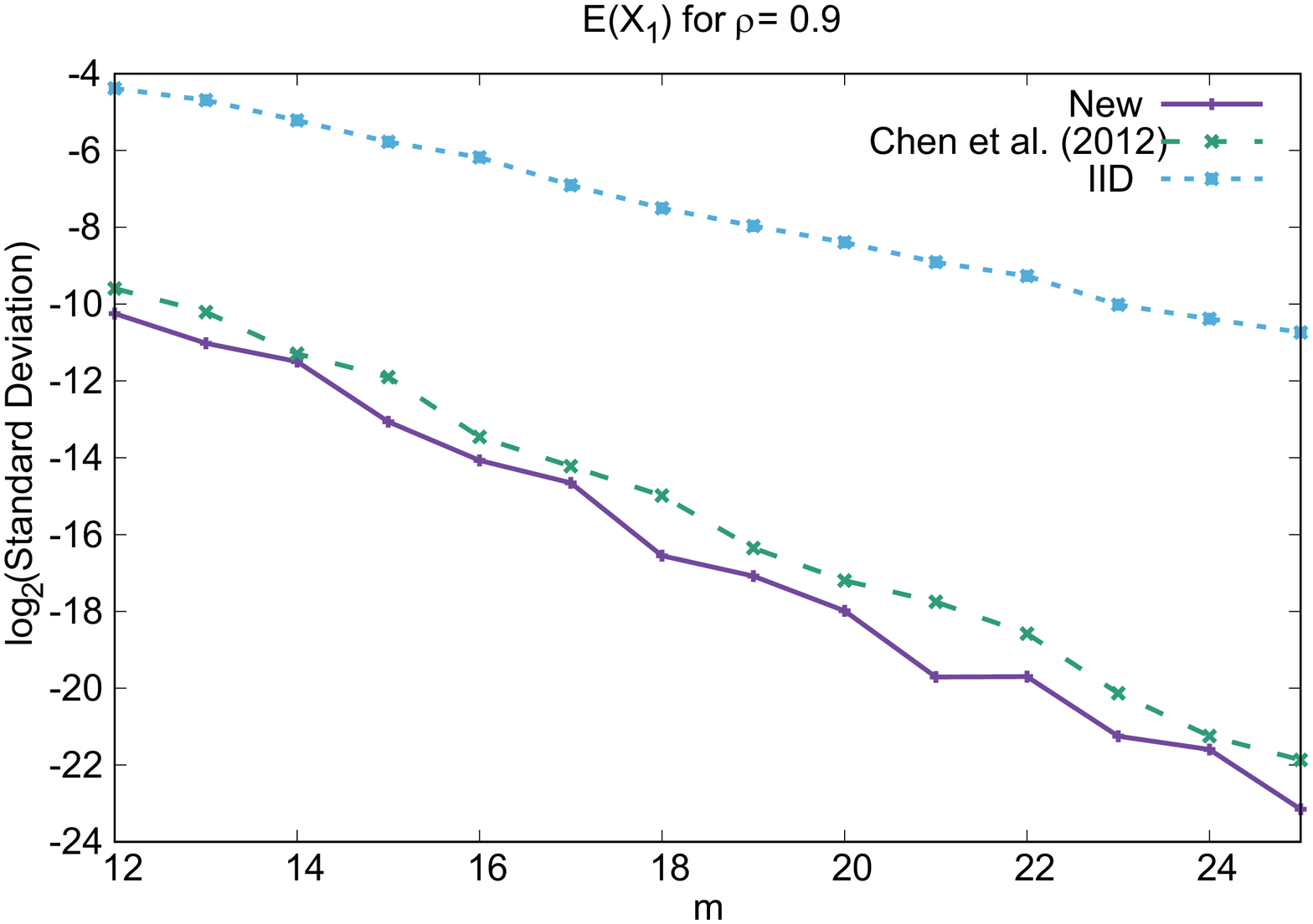}
\includegraphics[width=8.0cm]{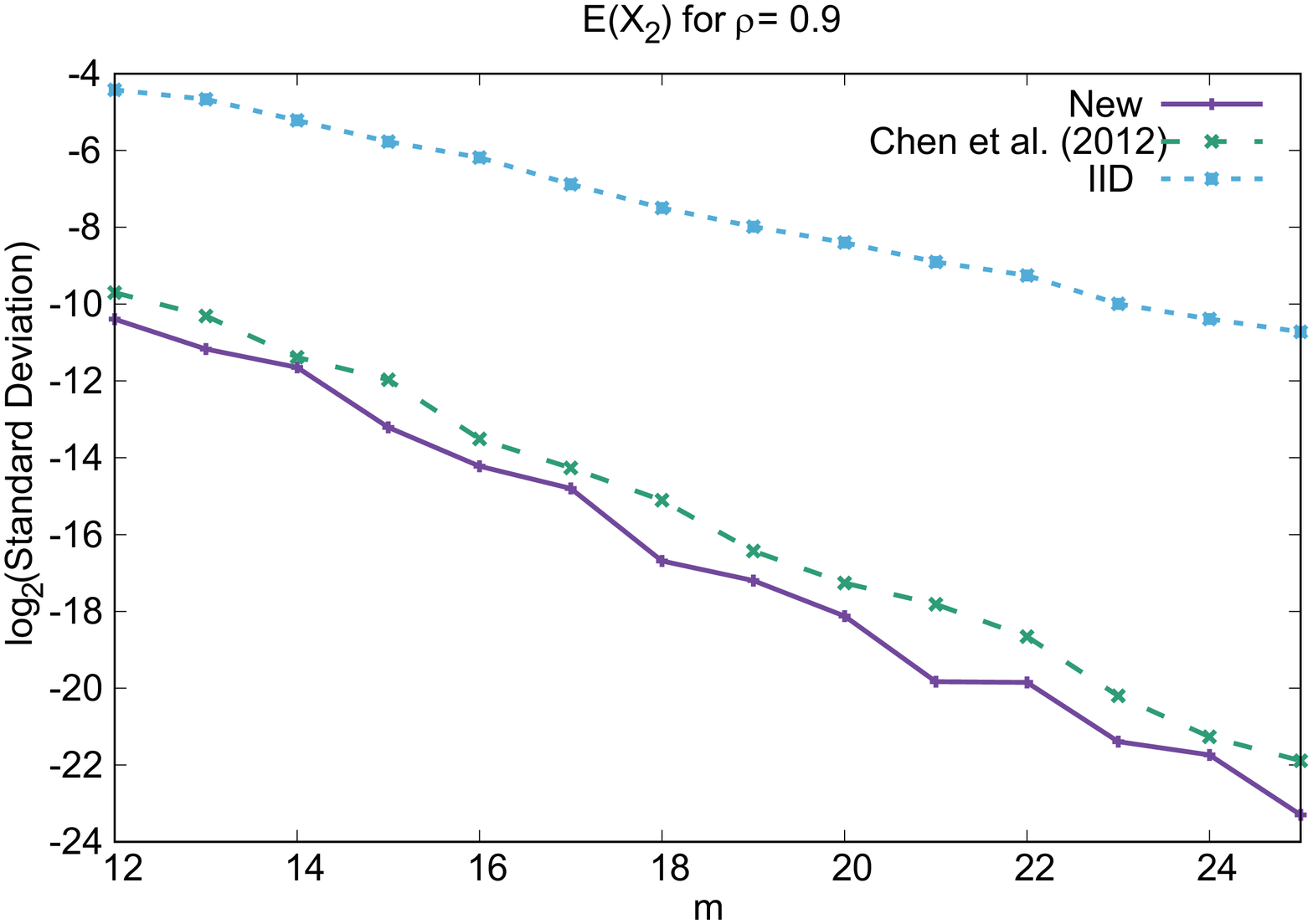}
\caption{Estimation of $E(X_1)$ and $E(X_2)$ for $\rho = 0, 0.3$ and $0.9$.}
\end{figure}

\begin{figure}[htbp] \label{fig:scatterplots}
\includegraphics[width=8.0cm]{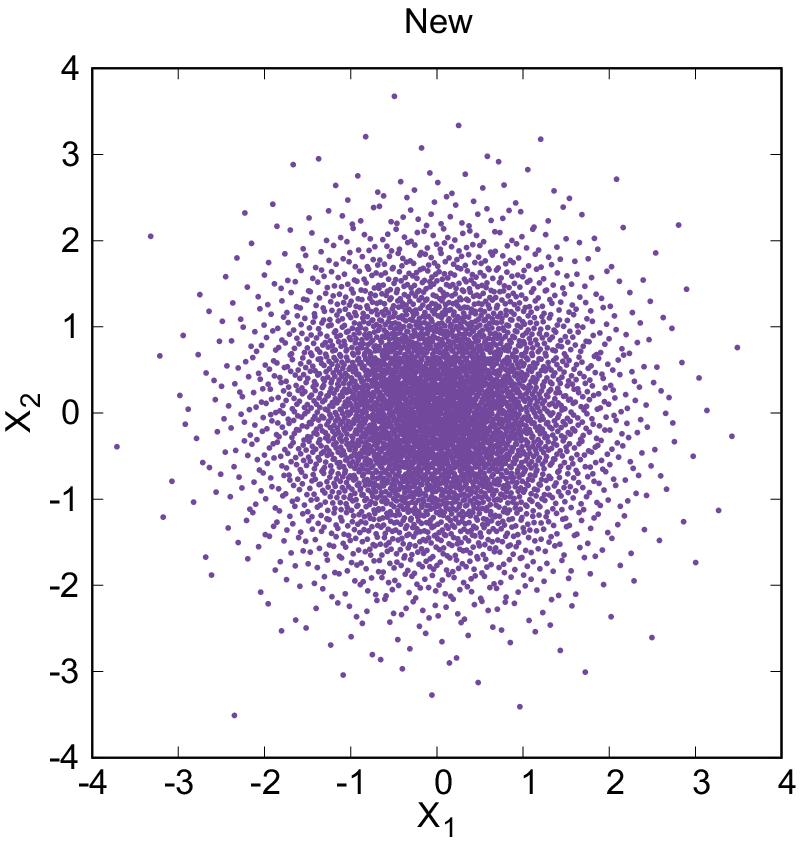}
\includegraphics[width=8.0cm]{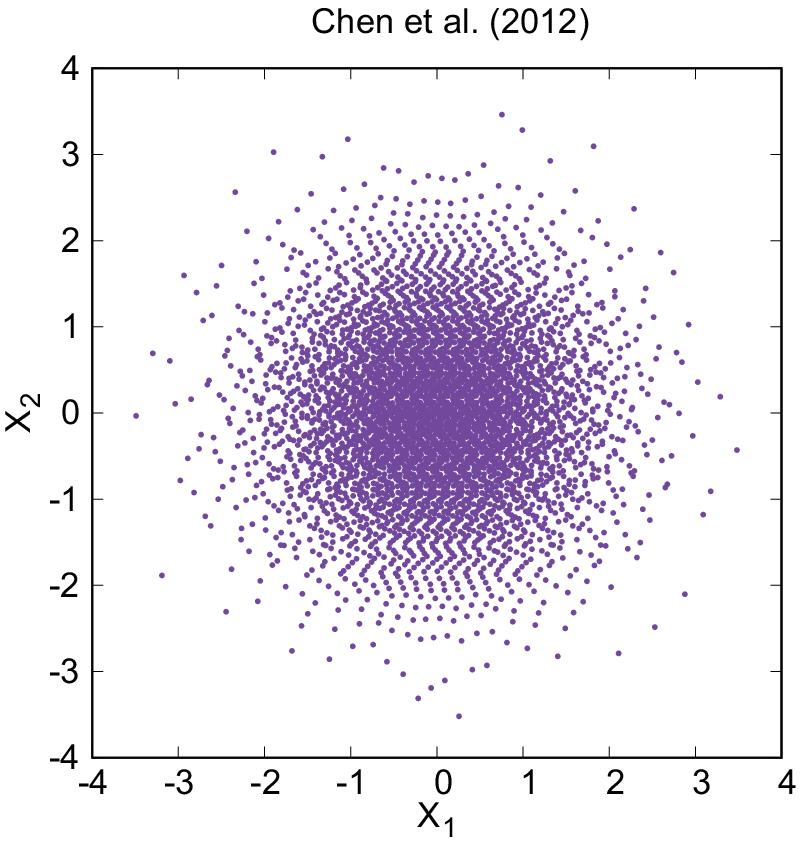}
\caption{Scatter plots of sampling $(X_1, X_2)$ for $\rho = 0$ and $m = 12$. }
\end{figure}

We estimate $E(X_1)$ and $E(X_2)$ 
by taking the sample mean. 
Hence, the true values are zero. 
We compare the following driving sequences:
\begin{enumerate}
\item New: our new Tausworthe generators;
\item Chen et al.~(2012): Tausworthe generators developed by Chen et al.~\cite{MR3173841}; and 
\item IID: Mersenne Twister \cite{Matsumoto:1998:MTE:272991.272995}. 
\end{enumerate}
Figure~\ref{fig:gaussian} shows a summary of standard deviations (in $\log_2$ scale) 
for $\rho = 0, 0.3$ and $0.9$ and $12 \leq m  \leq 25$ using $100$ digital shifts. 
Our new generators outperformed Chen's generators for no correlation $\rho = 0$ and weak correlation $\rho = 0.3$. 
Even for strong correlation $\rho = 0.9$, our new generators were still better than Chen's generators. 
In Figure~\ref{fig:scatterplots}, 
we generated scatter plots of sampling $(X_1, X_2)$ from 
our new and Chen's Tausworthe generators for $\rho = 0$ and $m = 12$. 
In the scatter plots, 
Chen's generator has a pattern of wiggly strips of points, 
which is optimized in terms of $64 \times 64$ grids for $s = 2$, 
but our generator seems to be highly balanced both for $X_1$ and $X_2$. 
Therefore, 
it can be expected that our new generators 
have better marginal distributions than the existing generators. 


\begin{figure}[htbp] \label{fig:gaussian2}
\begin{center}
\includegraphics[width=8.0cm]{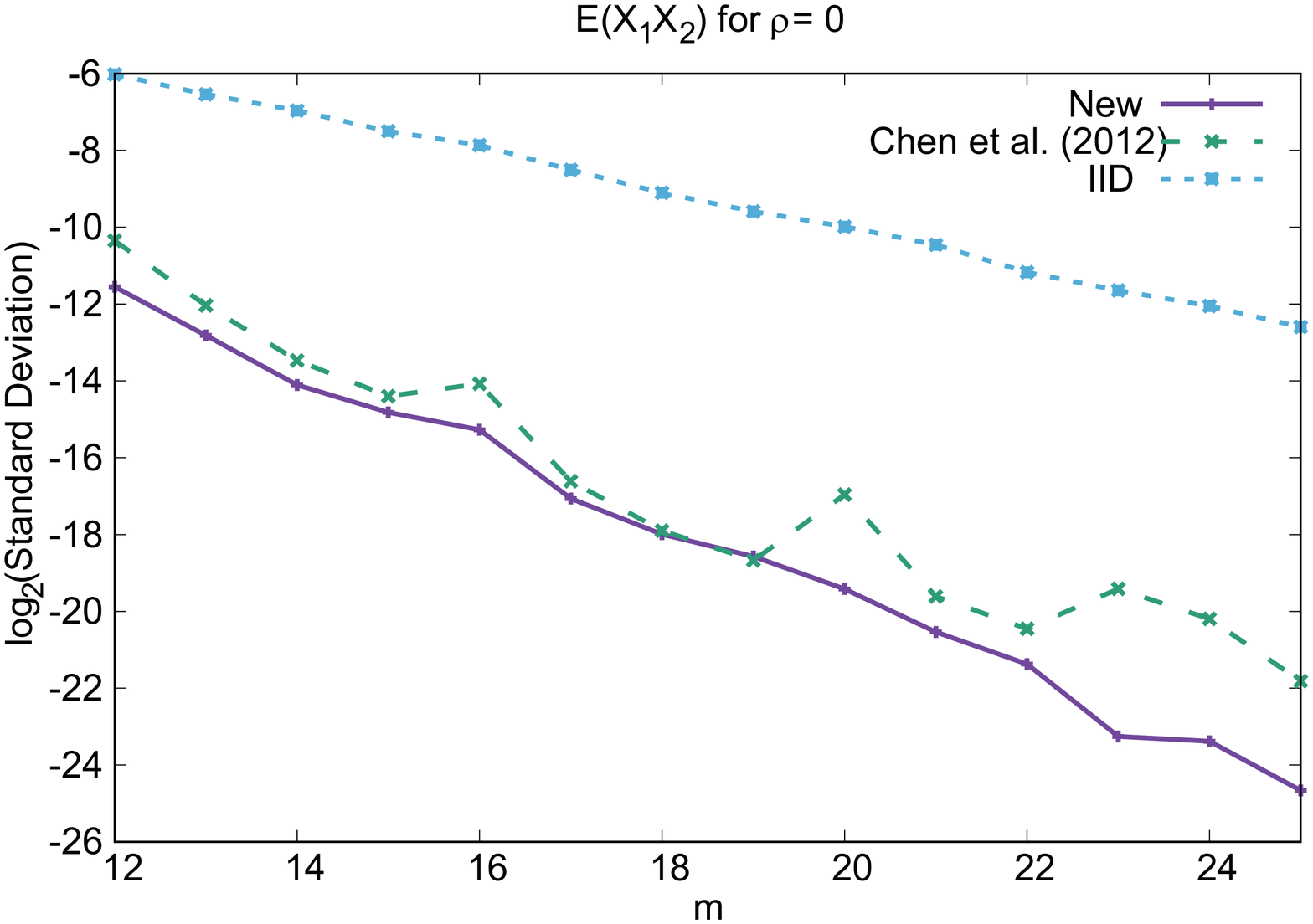}
\end{center}
\includegraphics[width=8.0cm]{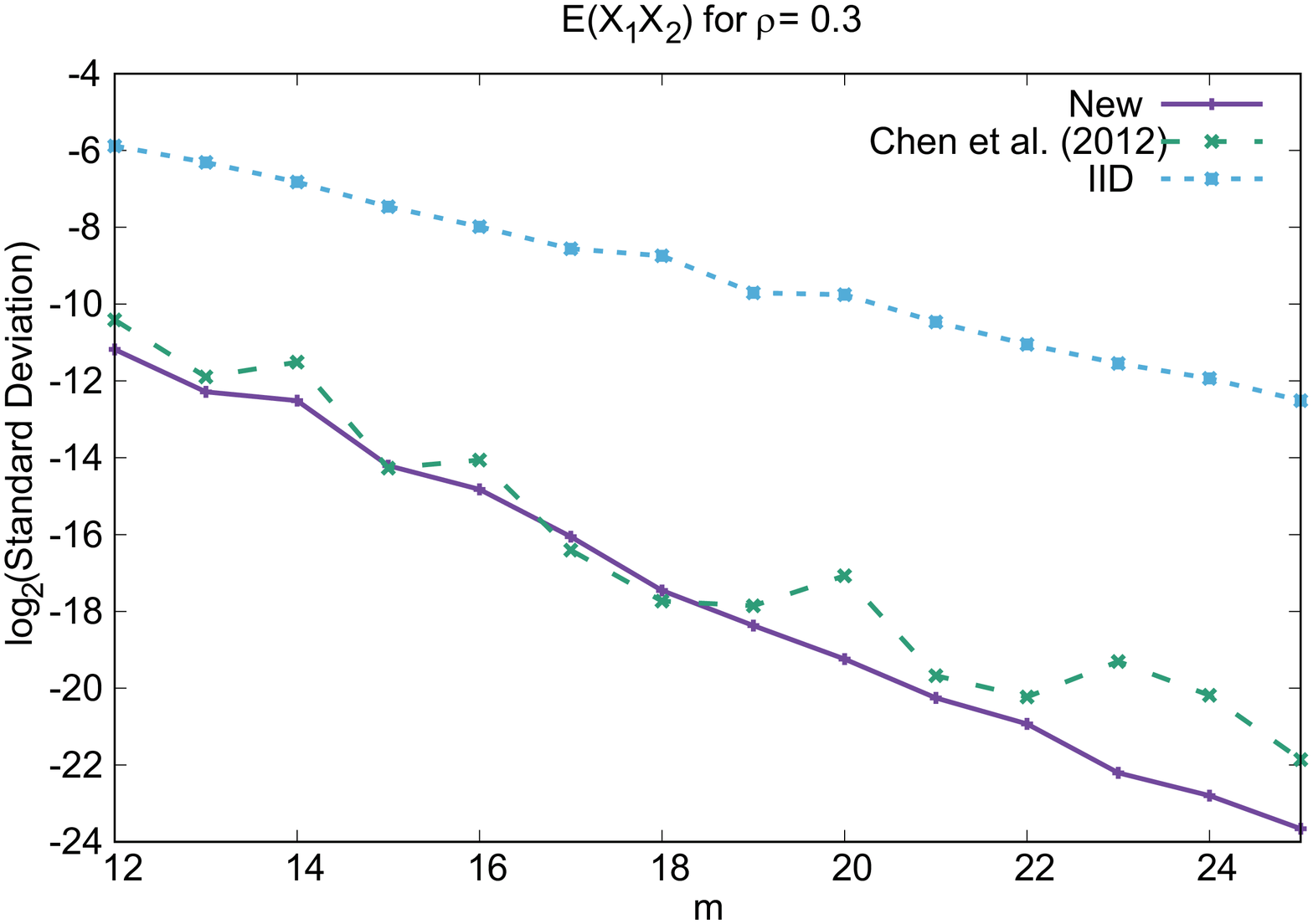}
\includegraphics[width=8.0cm]{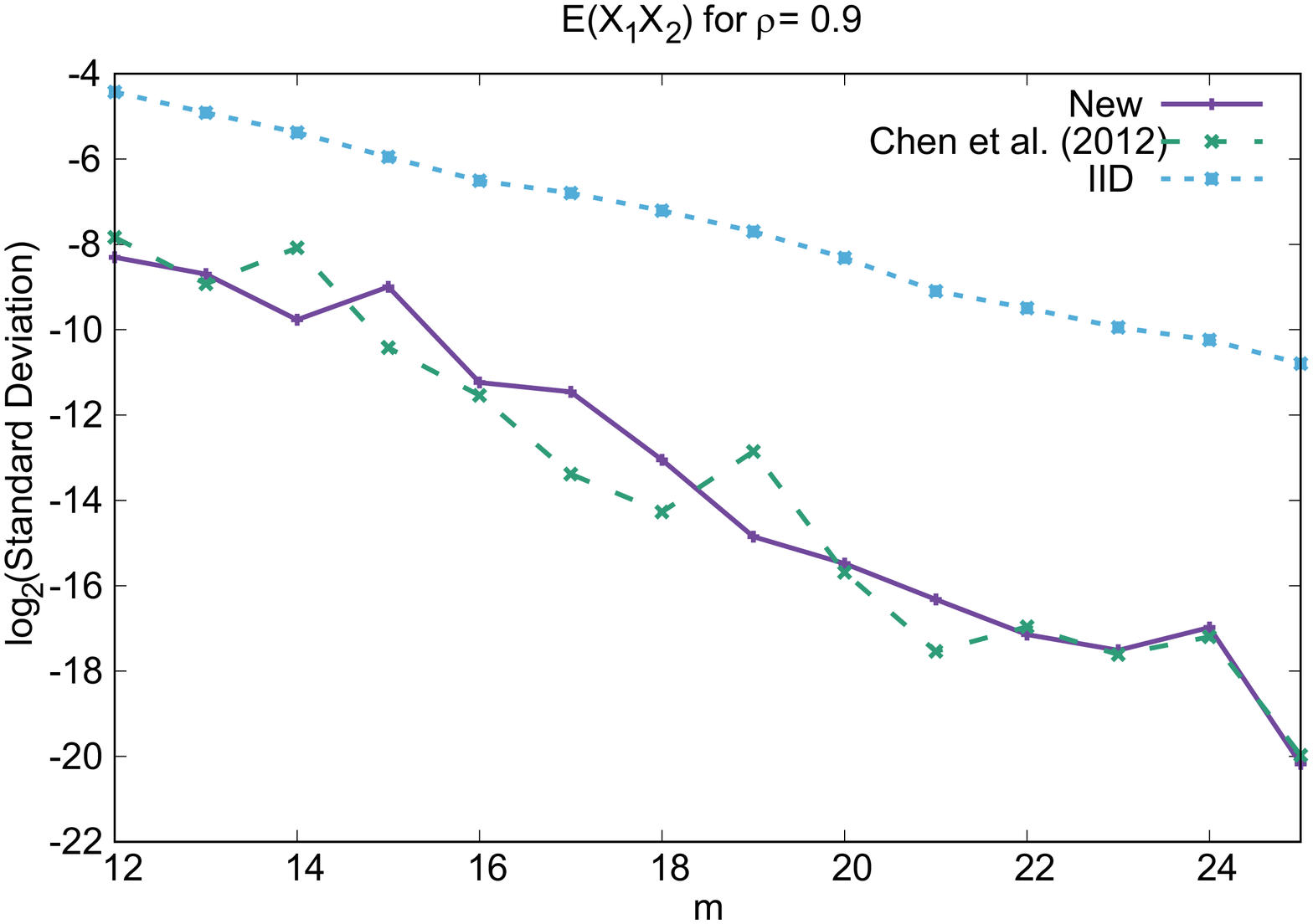}
\caption{Estimation of $E(X_1X_2)$ for $\rho = 0, 0.3$ and $0.9$.}
\end{figure}

In addition, as a test function, we estimated $E(X_1X_2)$, which has the true value $\rho$. 
Figure~\ref{fig:gaussian2} shows a summary of standard deviations (in $\log_2$ scale) 
for $\rho = 0, 0.3$ and $0.9$ and $12 \leq m  \leq 25$ using $100$ digital shifts. 
We obtained the results in which our new generators were superior to Chen's generators especially for $\rho = 0$ and $0.3$. 

\subsection{A hierarchical Bayesian model}

Our second example is a hierarchical Bayesian model \cite{MR1141740} 
used in \cite{MR2168266,MR2710331,doi:10.1080/10618600.1998.10474775}. 
Following \cite[Example~7.12]{MR2572239}, we explain the problem setting. 
We consider multiple failures of ten pumps in a nuclear plant, with the data given in Table~\ref{table:pump}. 
The modeling is based on the assumption that the number of failures 
of the $j$th pump follows a Poisson process with parameter $\lambda_j$ ($j = 1, \ldots, 10$). 
For an observation time $t_j$, 
the number of failures $X_j$ is thus a Poisson $\mathcal{P}(\lambda_j t_j)$ random variable. 
The standard prior distributions are gamma distributions $\mathcal{G}(\alpha, \beta)$ 
with shape parameter $\alpha$ and rate parameter $\beta$, 
which lead to the hierarchical model 
\begin{eqnarray*}
X_j & \sim & \mathcal{P}(\lambda_j t_j), \quad j = 1, \ldots, 10, \\
\lambda_j & \sim & \mathcal{G}(\alpha, \beta), \quad j = 1, \ldots, 10, \\
\beta & \sim & \mathcal{G}(\gamma, \delta),
\end{eqnarray*}
where the hyperparameter values are $\alpha = 1.802, \gamma = 0.1$, and $\delta = 1$. 
Our goal is to estimate the posterior means $E[\lambda_j]$ and $E[\beta]$ by taking the sample mean. 
For this purpose, we use a Gibbs sampler based on the full conditional distributions 
\begin{eqnarray*}
\lambda_j  \ |  \ \beta, t_j, x_j  & \sim & \mathcal{G}(x_j + \alpha, t_j + \beta), \quad j = 1, \ldots, 10, \\
\beta \ | \ \lambda_1, \ldots, \lambda_{10} & \sim &
\mathcal{G} \left( \gamma + 10 \alpha, \delta + \sum_{j = 1}^{10} \lambda_j \right).
\end{eqnarray*} 
Note that the state vector $(\lambda_1, \ldots, \lambda_{10}, \beta)$ has eleven dimensions. 
The starting point uses the maximum likelihood estimates $x_j/t_j$ for $\lambda_j$ 
together with 
the full conditional mean of $\beta$, given the starting $\lambda_j$. 
The Gibbs sampling is driven by inversion of gamma cumulative density functions. 
Similarly to (\ref{eqn:gibbs_output}), 
for the output values (\ref{eqn:Tausworthe sequence}), 
we define eleven-dimensional non-overlapping points $(u_0, \ldots, u_{10}), (u_{11}, \ldots, u_{21}), \ldots, (u_{11(N-2)}, \ldots, u_{11(N-1)-1})$, starting from the origin $(0, \ldots, 0)$, where $N= 2^m$ and $\gcd (2^m-1, 11)=1$. 

Table~\ref{table:variance} shows a summary of sample variances of posterior mean estimates 
for $m = 12, 14, 16$, and $18$ using 300 digital shifts. 
Our new Tausworthe generators were comparable to or even better than 
Chen's Tausworthe generators with a few exceptions (e.g., $\lambda_7, \lambda_8$, and $\lambda_9$ for $m = 14$). 
Such exceptions occurred in pumps for short monitoring periods, 
and this implies that it might be difficult to estimate those parameters with high accuracy  
from the perspective of Bayesian inference. 
In any case, our new generators were at least superior to IID uniform random number sequences generated by Mersenne Twister. 

\begin{table}[htbp] 
\caption{Number of failures and times of observation of ten pumps in a nuclear plant \cite{MR876882}.} \label{table:pump}
{\small
\begin{tabular}{c|rrrrrrrrrr} \hline
Pump $j$ & 1 & 2 & 3 & 4 & 5 & 6 & 7 & 8 & 9 & 10 \\ \hline
Failures $x_j$ & 5 & 1 & 5 & 14 & 3 & 19 & 1 & 1 & 4 & 22 \\ 
Time $t_j$ &  94.32 & 15.72 & 62.88 & 125.76 & 5.24 & 31.44  & 1.05 & 1.05 & 2.10 & 10.48 \\ \hline
\end{tabular}}
\end{table}

\begin{table}[htbp] 
\caption{Variance of posterior mean estimates for pump failure data.} \label{table:variance}
\begin{center}
{\small
\begin{tabular}{|c|ccccc|} \hline
\multicolumn{6}{|c|}{$m = 12$}\\ \hline
Parameter & $\lambda_1$ & $\lambda_2$ & $\lambda_3$ & $\lambda_4$ & $\lambda_5$\\ \hline
IID & 1.77e-07 & 1.98e-06 & 4.12e-07 & 1.96e-07 & 2.40e-05 \\
Chen & 4.77e-11 & 7.18e-10 & 8.91e-11 & 4.69e-11 & 7.44e-09 \\
New &  \textbf{8.13e-12} & \textbf{2.41e-10} & \textbf{1.96e-11} & \textbf{9.86e-12} & \textbf{4.11e-09} \\ \hline
\end{tabular}
\begin{tabular}{|c|cccccc|} \hline
Parameter & $\lambda_6$ & $\lambda_7$ & $\lambda_8$ & $\lambda_9$ & $\lambda_{10}$ & $\beta$ \\ \hline 
IID & 4.14e-06 & 9.79e-05 & 9.00e-05 & 1.05e-04 & 4.80e-05 & 2.29e-04 \\
Chen & 1.09e-09 & \textbf{1.03e-07} & 4.53e-08 & 3.81e-08 & 1.23e-08 & 1.68e-07 \\
New  &  \textbf{2.44e-10} & 1.78e-07 & \textbf{3.49e-08} & \textbf{2.38e-08} & \textbf{2.81e-09} & \textbf{5.21e-08} \\ \hline
\end{tabular}
\\
\begin{tabular}{|c|ccccc|} \hline
\multicolumn{6}{|c|}{$m = 14$}\\ \hline
Parameter & $\lambda_1$ & $\lambda_2$ & $\lambda_3$ & $\lambda_4$ & $\lambda_5$\\ \hline
IID & 4.33e-08 & 5.44e-07 & 9.21e-08 & 6.67e-08 & 5.46e-06 \\
Chen &  4.86e-12 & 1.07e-10 & 1.05e-11 & 5.15e-12 & 5.64e-09 \\
New &  \textbf{5.96e-13} & \textbf{2.48e-11} & \textbf{1.16e-12} & \textbf{6.13e-13} & \textbf{1.03e-09} \\ \hline
\end{tabular}
\begin{tabular}{|c|cccccc|} \hline
Parameter & $\lambda_6$ & $\lambda_7$ & $\lambda_8$ & $\lambda_9$ & $\lambda_{10}$ & $\beta$ \\ \hline 
IID & 1.12e-06 & 2.21e-05 & 2.32e-05 & 2.40e-05 & 1.21e-05 & 6.28e-05 \\
Chen & 9.75e-11 & \textbf{7.08e-09} & \textbf{1.37e-08} & \textbf{5.68e-09} & 1.46e-09 & 2.41e-08 \\
New  & \textbf{2.12e-11} & 5.60e-08 & 2.37e-07 & 1.18e-08 & \textbf{4.51e-10} & \textbf{4.86e-09} \\ \hline
\end{tabular}
\\
\begin{tabular}{|c|ccccc|} \hline
\multicolumn{6}{|c|}{$m = 16$}\\ \hline
Parameter & $\lambda_1$ & $\lambda_2$ & $\lambda_3$ & $\lambda_4$ & $\lambda_5$\\ \hline
IID & 1.08e-08 & 1.42e-07 & 2.42e-08 & 1.21e-08 & 1.44e-06 \\
Chen &  4.03e-13 & 8.28e-12 & 1.07e-12 & 4.73e-13 & 7.05e-11 \\
New &  \textbf{2.78e-14} & \textbf{1.53e-12} & \textbf{5.23e-14} & \textbf{2.40e-14} & \textbf{7.03e-11} \\ \hline
\end{tabular}
\begin{tabular}{|c|cccccc|} \hline
Parameter & $\lambda_6$ & $\lambda_7$ & $\lambda_8$ & $\lambda_9$ & $\lambda_{10}$ & $\beta$ \\ \hline 
IID & 3.01e-07 & 5.34e-06 & 5.65e-06 & 6.79e-06 & 2.61e-06 & 1.67e-05 \\
Chen &  8.74e-12 & 5.11e-10 & 5.34e-10 & 3.90e-10 & 9.90e-11 & 2.20e-09 \\
New  & \textbf{1.52e-12} & \textbf{2.16e-10} & \textbf{5.23e-10} & \textbf{1.71e-10} & \textbf{2.22e-11} & \textbf{1.05e-09} \\ \hline
\end{tabular}\\
\begin{tabular}{|c|ccccc|} \hline
\multicolumn{6}{|c|}{$m = 18$}\\ \hline
Parameter & $\lambda_1$ & $\lambda_2$ & $\lambda_3$ & $\lambda_4$ & $\lambda_5$\\ \hline
IID &  2.48e-09 & 3.21e-08 & 7.49e-09 & 3.47e-09 & 3.88e-07 \\
Chen & 2.50e-14 & 1.05e-12 & 5.12e-14 & 2.58e-14 & 1.86e-11 \\
New & \textbf{2.41e-15} & \textbf{7.97e-14} & \textbf{5.60e-15} & \textbf{1.86e-15} & \textbf{1.72e-12}  \\ \hline
\end{tabular}
\begin{tabular}{|c|cccccc|} \hline
Parameter & $\lambda_6$ & $\lambda_7$ & $\lambda_8$ & $\lambda_9$ & $\lambda_{10}$ & $\beta$ \\ \hline 
IID &  8.30e-08 & 1.34e-06 & 1.64e-06 & 1.65e-06 & 6.65e-07 & 4.24e-06 \\
Chen & 1.39e-12 & 7.52e-11 & 1.83e-10 & 9.83e-11 & 1.68e-11 & 9.01e-10 \\
New  & \textbf{7.84e-14} & \textbf{4.23e-11} & \textbf{8.99e-11} & \textbf{4.72e-11} & \textbf{3.80e-12} & \textbf{1.15e-10} \\ \hline
\end{tabular}}
\end{center}
\end{table}


\begin{remark}
In our experiments, we set $w = 32$. 
In fact, Chen et al.~\cite{MR3173841} 
originally defined Tausworthe generators in (\ref{eqn:Tausworthe}) 
with $m$-bit precision, that is, $u_i = \sum_{j = 0}^{m-1} a_{i \sigma +j} 2^{-j-1} \in [0,1)$. 
In this definition, we could not observe clear differences between our new generators and Chen's generators. 
However, we increased the precision of points and redefined Tausworthe generators with $w$ bits 
as in (\ref{eqn:Tausworthe}), and then the differences became clear. 
\end{remark}

\begin{remark} \label{remark:SMC}
Sequential Monte Carlo (SMC) 
can be used to perform Bayesian inference 
when the data are accumulated sequentially rather than being given a priori. 
Recently, Gerber and Chopin \cite{MR3351446} developed a class of algorithms 
combining SMC and randomized QMC to accelerate convergence.
\end{remark}

\section{Conclusion} \label{sec:conclusion}

We conducted an exhaustive search of short-period Tausworthe generators 
for Markov chain QMC in terms of the $t$-value. 
Our key technique was to use the continued fraction expansion of $q(x)/p(x)$ 
by refining the algorithm of 
Tezuka and Fushimi \cite{MR1160278} on modern computers. 
As a result, we obtained the point sets 
with $t$-values optimal for $s = 2$ and small for $s \geq 3$. 
We also reported numerical examples using Gibbs sampling in which  
our new generators performed better than the existing generators of Chen et al.~\cite{MR3173841}. 
The code in C is available at \url{https://github.com/sharase/cud}.

As a future work, we will attempt more realistic numerical examples 
as in \cite{ChenThesis,MR2426105,MR2710331}. 
For this purpose, we believe that the next task is to embed 
our new and existing generators into several programming languages 
for statistical computing; for example, R, Stan, and Python. 
Thus, we are now planning a software implementation of Markov chain QMC. 




\section*{Acknowledgements}
This work was supported by JSPS KAKENHI Grant Numbers JP18K18016, JP26730015, JP26310211, JP15K13460. 
The author would like to thank the anonymous reviewers for many valuable comments and suggestions. 

\bibliography{harase-cud}





\end{document}